\documentclass[twocolumn,english,aps,prb,showpacs,superscriptaddress,amssymb,amsfonts]{revtex4}
\usepackage[T1]{fontenc}
\usepackage[latin9]{inputenc}
\usepackage{amsmath}
\usepackage{braket}
\usepackage{epstopdf}
\usepackage{bm}
\usepackage{graphicx}
\usepackage{amssymb}
\usepackage{color}
\usepackage{enumerate}
\usepackage[normalem]{ulem}

\makeatletter
\newcommand{\be}{\begin{equation}}
\newcommand{\ee}{\end{equation}}                  
\newcommand{\bea}{\begin{eqnarray}}
\newcommand{\eea}{\end{eqnarray}}
\newcommand{\beas}{\begin{eqnarray*}}
\newcommand{\eeas}{\end{eqnarray*}}

\newcommand{\tr}{\textrm{tr}}
\newcommand{\Abar}{\overline{A}}
\newcommand{\Vabar}{V_{\overline{A}}}

\makeatother

\usepackage{babel}

\begin{document}
 
\title{Does a single eigenstate encode the full Hamiltonian?}

\author{James R. Garrison} 
\affiliation{Department of Physics, University of California, Santa Barbara, California 93106, USA}

\author{Tarun Grover}
\affiliation{Kavli Institute for Theoretical Physics, University of California, Santa Barbara, CA 93106, USA}

\begin{abstract}
The Eigenstate Thermalization Hypothesis (ETH) posits that the reduced density matrix for a subsystem corresponding to an excited eigenstate is ``thermal.'' Here we expound on this hypothesis by asking: for which class of operators, local or non-local, is ETH satisfied? We show that this question is directly related to a seemingly unrelated question: is the Hamiltonian of a system encoded within a single eigenstate?  We formulate a strong form of ETH where in the thermodynamic limit, the reduced density matrix of a subsystem corresponding to a pure, finite energy density eigenstate asymptotically becomes equal to the thermal reduced density matrix, as long as the subsystem size is much less than the total system size, irrespective of how large the subsystem is compared to any intrinsic length scale of the system. This allows one to access the properties of the underlying Hamiltonian at arbitrary energy densities/temperatures using just a \textit{single} eigenstate. We provide support for our conjecture by performing an exact diagonalization study of a non-integrable 1D lattice quantum model with only energy conservation. In addition, we examine the case in which the subsystem size is a finite fraction of the total system size, and find that even in this case, a large class of operators  continue to match their canonical expectation values. Specifically, the von Neumann entanglement entropy equals the thermal entropy as long as the subsystem is less than half the total system. We also study, both analytically and numerically, a particle number conserving model at infinite temperature which substantiates our conjectures.
\end{abstract}

\maketitle
\tableofcontents
\section{Introduction} \label{sec:intro}

Given a local Hamiltonian, what information about the system is encoded in a single eigenstate? If the eigenstate happens to be a ground state of the Hamiltonian, tremendous amount of progress can be made on this question for Lorentz invariant systems \cite{bisognano1976, susskind2004}, especially conformal field theories (CFTs) \cite{holzhey1994, callan1994, calabrese2004, casini2011}, and for topological phases \cite{levin2006, kitaev2006, li2008}.  For example, one can read off the central charge of a CFT from the ground state entanglement \cite{holzhey1994, callan1994, calabrese2004}, while for topological phases, essentially all `topological data' such as braiding statistics of anyons can be extracted from the degenerate ground states \cite{kitaev2006, li2008, zhang2012}. In this paper we argue that a single finite energy density eigenstate of an ergodic quantum many-body Hamiltonian is sufficient to determine the properties of the system at all temperatures.

It is not very surprising that the \textit{ground states} of quantum many-body systems contain some information about their excitations. This is because an entanglement cut often mimics an actual physical cut through the system, thus exposing the underlying excitations along the entangling boundary \cite{li2008}. The same intuition is tied to the fact that the ground state entanglement satisfies a ``boundary law'' of entanglement entropy \cite{bombelli1986,srednicki1993}, that is, the von Neumann entanglement entropy $S_1 = -\tr_A\left(\rho_A \log(\rho_A)\right)$ of the ground state corresponding to a subsystem $A$ scales with the size of the boundary of subsystem $A$.

How does the nature of information encoded evolve as one goes from the ground state to an excited eigenstate? Typically, there always exist eigenstates with energy $E$ just above the ground state which continue to satisfy an area law of entanglement. These are the eigenstates which have zero energy density, i.e.\ $\lim_{V \to \infty} \frac{E-E_0}{V} =0$ where $E_0$ is the ground state energy and $V$ is the total volume of the system. These eigenstates can often be interpreted as the action of a sum of local operators acting on the ground state; for example, in a system with spontaneous symmetry breaking one can construct an eigenstate consisting of a few magnons by a superposition of spin-flips acting on the ground state. Furthermore, the level spacing between two contiguous low-lying excitations scales as $\delta E \sim 1/L^{\alpha}$ where $\alpha >0$ depends on dimensionality and the phase of matter under consideration. In this paper, we will instead be concerned with excited eigenstates that have a \textit{finite energy density}, i.e.\ $\lim_{V \to\infty} \frac{E-E_0}{V} \neq 0$. For notational convenience, we will set $E_0=0$ for the remainder of this paper.

As argued by Srednicki \cite{srednicki1994}, a typical finite energy density state (i.e.\ a typical state in the Hilbert space that satisfies $\langle \psi|H|\psi\rangle =  V e$ where $e$ is the energy density) when time-evolved with the Hamiltonian $H$ for sufficient time is expected to lead to predictions dictated by the basic tenets of equilibrium statistical mechanics, \textit{if} the system thermalizes.  Such an expectation leads to the ``Eigenstate Thermalization Hypothesis'' (ETH)  \cite{deutsch1991, srednicki1994, srednicki1998}, which stipulates that the thermalization occurs at the level of each individual eigenstate. An alternative approach by Deutsch \cite{deutsch1991}, which is based on perturbing an integrable system by a small integrability breaking term, leads to the same suggestion. If ETH holds true, then in the thermodynamic limit the equal-time correlators of an operator with respect to a finite energy density eigenstate $|\psi\rangle$ are precisely equal to those derived from a thermal ensemble, i.e.

\be 
\langle \psi| O |\psi \rangle = \frac{\tr \left(\,O e^{-\beta H}\right)}{\tr \left(\,e^{-\beta H}\right)}  \label{eq:eth1}
\ee
where $\beta$ is chosen such that the Eq.\ \ref{eq:eth1} holds true when $O = H$, the Hamiltonian. Henceforth we will use the notation $|\psi\rangle_\beta$ to denote an eigenstate whose energy density corresponds to temperature $\beta^{-1}$. A notable exception to ETH is a many-body localized system in the context of strongly disordered interacting quantum systems, \cite{gefen1997, gornyi2005,basko2006,huse2007,bela, imbrie, nandkishore2014} which fails to thermalize and does not satisfy Eq.\ \ref{eq:eth1}. The possibility\cite{kagan1, kagan2, grover2014, muller_aps, muller2014, hickey, yao2014}, or impossibility\cite{roeck1, roeck2, roeck3, abanin2015}, of the violation of ETH without disorder has also been discussed recently.

In this paper, we restrict ourselves to systems where ETH, as defined by Eq.\ \ref{eq:eth1}, holds. However, Eq.\ \ref{eq:eth1} alone is incomplete unless one also specifies the class of operators for which it holds.  For example, one simple non-local operator for which Eq.\ \ref{eq:eth1}  breaks down is the projection operator $|\psi\rangle\langle\psi|$ onto the eigenstate $|\psi\rangle$ that enters Eq.\ \ref{eq:eth1}; the left hand side of Eq.\ \ref{eq:eth1} yields unity for this operator, while the right hand side is exponentially small in the volume, a clear disagreement.  On that note, it is often mentioned that in systems where Eq.\ \ref{eq:eth1} does hold, it does so only for ``few body'' operators \cite{rigol2008, lauchli2011, srednicki2011} where, to our knowledge, the precise meaning of few-body operator has not been  clarified (see Ref.\ \onlinecite{hosur} for related discussion).  In this paper, we conjecture and provide numerical evidence that Eq.\ \ref{eq:eth1} holds for \textit{all} operators within a subsystem $A$ when the volume $V_A$ of subsystem $A$ satisfies $V_A \ll V$ (or, more precisely, when $V_A / V \rightarrow 0$ as $V \rightarrow \infty$).  We also explore the more general case where subsystem $A$ spans a finite fraction $f\equiv V_A/V >0$ of the total system size.  We provide some evidence that when the fraction is less than a critical $O(1)$ number $f^{*}$, then all operators \emph{not} explicitly involving energy conservation take their thermal values.  We also explore the more general condition $V_A < V/2$ and show that even in this case, Eq.\ \ref{eq:eth1} holds for a large class of operators. On that note, we should mention that the  questions such as which Hamiltonians (and which operators) satisfy ETH is now entering the realm of experimental physics (see e.g.\ Ref.\ \onlinecite{bloch}) due to advances in high resolution imaging techniques \cite{greiner2009}.

The satisfaction of Eq.\ \ref{eq:eth1} for all operators in a subsystem $A$ is equivalent to the statement that the reduced density matrix $\rho_A(|\psi\rangle_{\beta}) = \tr_{\overline{A}} |\psi\rangle_{\beta} {}_{\beta}\langle \psi|$ corresponding to an eigenstate $|\psi\rangle_{\beta}$ is given by

\be 
\rho_A(|\psi\rangle_{\beta}) = \rho_{A,\mathrm{th}}(\beta) \tag{2a} \stepcounter{equation} \label{eq:eth2}
\ee
where 
\be 
\rho_{A,\mathrm{th}}(\beta) =  \frac{\tr_{\overline{A}} \left( e^{-\beta H}\right)}{\tr \left(e^{-\beta H}\right)}, \nonumber
\ee
$\Abar$ being the complement of $A$. Note that the trace in the denominator is over the whole Hilbert space.  When $V_A$ is held constant, the equality in Eq.\ \ref{eq:eth2} means the density matrices become elementwise equal in any basis as $V \rightarrow \infty$.

One immediate consequence of Eq.\ \ref{eq:eth2} is that the  thermodynamical properties of a system at arbitrary temperatures can be calculated using a \textit{single} eigenstate. For example, Eq.\ \ref{eq:eth2} implies that to the leading order, the Renyi entropies $S_\alpha$ ($= - \frac{1}{\alpha - 1} \log \left[ \tr_A (\rho_A^\alpha) \right]$) for an eigenstate $|\psi\rangle_{\beta}$ corresponding to a subsystem $A$  with $V_A \ll V$ are given by

\be 
S_\alpha =  \frac{\alpha}{\alpha-1} V_A \beta \left( f(\alpha\beta) - f(\beta) \right), \label{eq:Sn}
\ee
where $f(\beta)$ is the free energy density at temperature $\beta^{-1}$.  The above equation allows one to access the free energy density $f$ at an arbitrary temperature by varying $\alpha$. Note that Eq.\ \ref{eq:Sn} holds only to the leading order because Renyi entropies $S_\alpha$ receive additional subleading contributions due to the conical singularity induced at the boundary of subsystem $A$ \cite{holzhey1994, callan1994, calabrese2004}.  In the limit $\alpha \rightarrow 1$, one recovers the equality between the von Neumann entanglement entropy $S_1$ and the thermal entropy $S_\mathrm{th} = V_A s_\mathrm{th}(\beta)$, where $s_\mathrm{th}(\beta)$ is the thermal entropy density at temperature $\beta^{-1}$, a result which was argued to hold in Ref.\ \onlinecite{deutsch2010} for the special case of two weakly coupled ergodic systems. We emphasize that these results cannot be derived from Eq.\ \ref{eq:eth1} alone were it to hold only for local operators, since entanglement entropies do not correspond to the expectation value of any local operator. We also note that Refs.\ \onlinecite{lauchli2013, melko2011} simulated the thermal Renyi entropy $S_\alpha$ (starting with the expression on the right hand side of Eq.\ \ref{eq:eth2})  using Quantum Monte Carlo to access the properties of the system at temperature $(\alpha\beta)^{-1}$. Of course, Quantum Monte Carlo methods are not well suited to verifying ETH since they cannot access properties of a \textit{single} eigenstate  (the left hand side of Eq.\ \ref{eq:eth2}).

We will also discuss an approximate, but more intuitive form of ETH, given by
\be
\rho_A(|\psi\rangle_{\beta}) \approx \frac{e^{-\beta H_A}}{\tr_A \left(e^{-\beta H_A}\right)} \tag{2b} \label{eq:eth3}
\ee
where $H_A$ is the projection of the original Hamiltonian onto subsystem $A$. This form is approximate compared to Eq.\ \ref{eq:eth2} because generically, it does not capture the correlations near the boundary correctly due to the somewhat arbitrary truncation scheme used to obtain $H_A$. Nevertheless, equations \ref{eq:eth2} and \ref{eq:eth3} both yield the same results for all bulk quantities such as the Renyi entropy densities, as well as correlation functions of operators that have support only far from the boundary.

A central task of this paper is to check the validity of Eqs.\ \ref{eq:eth2} and \ref{eq:eth3} and their consequences for model non-integrable systems. As already mentioned, we will argue that ETH allows one to calculate thermodynamical quantities as well as correlators \textit{at all temperatures/energy densities using only a single eigenstate}. We will demonstrate this explicitly by studying a quantum 1D model numerically. 

As mentioned above, we find evidence that Eq.\ \ref{eq:eth1} holds for many operators even when $V_A/V$ is held constant with $V_A/V $ less than some number $f^{*} > 0$. In particular, as we discuss later, our results strongly indicate that $f < 1/2$ is sufficient to guarantee equivalence between the von Neumann entropy density of a pure eigenstate, and the thermal entropy density at the corresponding temperature. This is in contrast to Ref.\ \onlinecite{santos2012} where it was argued that such an  equivalence holds only in the limit $f^{*} \rightarrow 0$. Recently \cite{storms2014, lai2014}, the requirement $f^{*} \rightarrow 0$ was substantiated using analytical and large scale numerical calculations for  \textit{free} fermions, an integrable system. Our results indicate that the $f^{*} \rightarrow 0$ requirement is likely a consequence of the integrable nature of the models in Refs.\ \onlinecite{storms2014, lai2014}.

The paper is organized as follows.  Sec.\ \ref{sec:general} discusses general considerations for the validity of ETH, and introduces a division of all operators in a given subsystem into two distinct classes, which have different requirements for ETH to hold.  Sec.\ \ref{sec:mumodel} illustrates some general features of ETH by studying properties of a hardcore boson model with global particle number conservation for infinite temperature eigenstates.  Sec.\ \ref{sec:tfi} introduces the model we study in the remainder of the paper, the transverse field Ising model with longitudinal field.  Sec.\ \ref{sec:Sn} focuses on the entanglement entropies at finite temperature. Sec.\ \ref{sec:extracting} studies the validity of ETH when $V_A \ll V$ by providing a close look into the entanglement Hamiltonian, focusing on its spectrum and Schmidt vectors.  This section also demonstrates the validity of Eq.\ \ref{eq:eth2} when $V_A \ll V$ by considering the trace norm distance between both sides.  Sec.\ \ref{sec:results_finite_ratio} explores the validity of ETH when $V_A/V$ is taken to be finite as $V \rightarrow \infty$. Sec.\ \ref{sec:corr} provides an application, by using the reduced density matrix from a single eigenstate to predict correlators at all (finite) temperatures.  Sec.\ \ref{sec:discuss} summarizes our results and provides thoughts for future discussion.

\section{General considerations} \label{sec:general}

\subsection{Determining the Hamiltonian from microstates in classical statistical mechanics}  \label{sec:classical}
Suppose, for an isolated system described by classical statistical mechanics in a total volume $V$, we are given access to all classical microstates in a small energy window $[E,E+\Delta E]$, where  $\Delta E \sim \sqrt{V}$ is on the order of the energy fluctuations in the total system were the system coupled to a thermal bath, and thus all microstates correspond to the same \textit{energy density}. We pose the question: does this information suffice to determine the underlying Hamiltonian, assuming that the Hamiltonian is local? The answer is indeed yes, following the standard procedure of obtaining the canonical ensemble from a microcanonical ensemble. In particular, let us make a fictitious division of the system into $A$ and $\overline{A}$ such that $V_A \ll V_{\overline{A}}$, and count the number of times a particular configuration $C_A$ appears in subsystem $A$. This determines the probability distribution for finding a given configuration, $P(C_A)$. If all microstates are equally likely, then \cite{pathria}
\be 
P(C_A) = \frac{e^{-\beta E (C_A)}}{\sum_{\{C_A\}} e^{-\beta E (C_A)}}  \label{eq:classicalpdf}
\ee
where $E (C_A)$ is the energy in subsystem $A$.  One may now invert this equation to obtain the energy $E(C_A) = -\frac{1}{\beta} \log(P(C_A))$, up to an irrelevant constant shift of energy. In a classical statistical mechanical system $E(C_A)$ \textit{is} the Hamiltonian for subsystem $A$. In particular, knowing $E(C_A)$, one may now calculate any thermodynamic property at $\textit{any}$ temperature. Here it is crucial to note that Eq.\ \ref{eq:classicalpdf} does not assume that the energy density $E(C_A)/V_A$ equals the energy density $E/V$ of the microstates being sampled.

As discussed in the introduction, we will provide evidence that the quantum mechanical analog of Eq.\ \ref{eq:classicalpdf} is given by Eqs.\ \ref{eq:eth2}, \ref{eq:eth3}. We now proceed to discuss the conditions under which Eqs.\ \ref{eq:eth2}, \ref{eq:eth3} are valid.

\subsection{Two classes of operators} \label{sec:twoclass}

For reasons soon to be discussed, we find it useful to separate operators in a given Hilbert space into two classes:

{\bf Class I (``Equithermal Operators''):} If the reduced density matrix takes the thermal form (i.e.\ the right hand side of Eq.\ \ref{eq:eth3}), then in the limit $V_A \rightarrow \infty$, the expectation value of equithermal operators  receives contribution only from the eigenstates of $H_A$ at an energy density corresponding to the temperature $\beta^{-1}$. One might have thought that this is true for all operators, however, there exist operators such as $e^{-n\beta H_A}$, whose expectation value includes contribution from eigenstates of $H_A$ at temperature $[(n+1)\beta]^{-1}$ in addition to the temperature $\beta^{-1}$. Clearly, local operators fall into this class, as do sums of local operators.  Several non-local operators, including the von Neumann entropy $S_1$, also fall into this class.

{\bf Class II (``Non-equithermal Operators''):}  We dub all operators not in Class I as ``non-equithermal operators'', or Class II operators. All Renyi entropies $S_\alpha$ (for $\alpha \neq 1$) fall into this class\cite{baez}.
\subsection{ ETH: Class I vs.\ Class II operators}  \label{sec:ethclass}

Let us first consider the relationship between Eq.\ \ref{eq:eth1} and Eqs.\ \ref{eq:eth2}, \ref{eq:eth3}. Eq.\ 1 may be rewritten as,

\be 
\tr_A\left( \rho_A O\right) =  \frac{\tr_A \left(\,O\, \tr_{\Abar} \left( e^{-\beta H}\right)\right)}{\tr \left(\,e^{-\beta H}\right)} \label{eq:eth4}
\ee
If this equation holds for \textit{all} operators in a subsystem $A$, hermitian as well as non-hermitian, then one obtains Eq.\ \ref{eq:eth2}, $\rho_A(|\psi\rangle_{\beta}) = \rho_{A,\mathrm{th}}(\beta)$. This is because one may expand both the $\rho_A$ and $\rho_{A,\mathrm{th}}$ in terms of the complete set of operators in subsystem $A$, and by choosing appropriate $O$ prove that they are equal to  each other element-by-element. One of the most important consequences of this equality is that it allows one to extract properties of the Hamiltonian at arbitrary temperatures using a single eigenstate, which is one of the central points of this paper.

We will now discuss ETH for both Class I and Class II operators.  For each class, we consider separately two cases: (i) when $V_A \ll V$; and (ii) when the ratio $f \equiv V_A/V$ is taken to be fixed and finite as $V_A, V \rightarrow \infty$.

\subsubsection{ETH for Class I operators}  \label{sec:classI}

Let us first briefly discuss ETH for Class I operators when $V_A \ll V$.  This includes both the case where $V_A$ is held fixed as $V \rightarrow \infty$, as well as the case where the limits $V_A, V \rightarrow \infty$ are taken such that $V_A/V \rightarrow 0$.  In fact, this is the traditional definition of ETH---that all local, ``few-body'' operators match their values in the canonical ensemble in this case.

Let us now consider the validity of ETH for Class I operators in the fixed-ratio limit where $0 < f < \frac12$ is finite.  In contrast to classical statistical mechanics, we expect that quantum mechanically, one does not require the constraint $V_A \ll \Vabar$ for ETH to hold for a large class of Class I operators. Indeed, as discussed below, several known results already point to the conclusion that Eq.\ \ref{eq:eth1} holds for at least some operators in Class I, as long as $V_A < \Vabar$ with both  $V_A, \Vabar \rightarrow \infty$.

One piece of evidence  that suggests that Eq.\ \ref{eq:eth1} might hold for Class I operators as long as $V_A < \Vabar$ comes from the study of quantum quenches in conformal field theories (CFTs). As shown in Ref.\ \onlinecite{cardy2014}, the time-dependent reduced density matrix $\rho_A(t)$ of a system initially prepared in a low-entanglement state, and evolved  with a CFT Hamiltonian, approaches the thermal density matrix, as long as $V_A < V/2$, with $V_A, V \rightarrow \infty$. Ref.\ \onlinecite{cardy2014} characterized the closeness between $\rho_A(t)$ and the thermal density matrix $\rho_{A,\mathrm{th}}$ (Eq.\ \ref{eq:eth2}) in terms of the operator overlap $I(t) = \frac{\tr(\rho_A(t) \rho_{A,\mathrm{th}})}{\left(\tr(\rho^2_A(t)) \tr(\rho^2_{A,\mathrm{th}})\right)^{1/2}}$, which is exponentially close to unity for $V_A/2 < t < \Vabar/2$. It is important to note that in the thermodynamic limit, $I$ only receives contribution from eigenstates at temperature $\beta^{-1}$, so this only guarantees that operators in Class I will satisfy Eq.\ \ref{eq:eth1}.

Another piece of evidence comes from the recent studies of large central charge conformal field theories \cite{asplund2014, caputa2014, kaplan2015}. In particular, Refs.\ \onlinecite{asplund2014, kaplan2015} studied the entanglement entropy of pure eigenstates in finite temperature conformal field theories with large central charge. In the limit $V_A, V \gg 1/T$, while keeping $V_A/V$ fixed, it was found that the entanglement entropy becomes equal to the thermal entropy at all non-zero temperatures as long as $V_A < \Vabar$.

Lastly, the entanglement entropy for a random pure state is given by \cite{lubkin1978, lloyd1988, page1993}:

\be 
S = -\log\left({|\mathcal{H}_A|}^{-1} + {|\mathcal{H}_{\overline{A}}|}^{-1} -{|\mathcal{H}|}^{-1}\right) \label{eq:EErand}
\ee
 where $|\mathcal{H}_A|, |\mathcal{H}_{\overline{A}}|, |\mathcal{H}|$ are the sizes of the Hilbert spaces of subsystems $A$, $\Abar$ and the total system ($= A \cup \Abar$) respectively. Thus, as soon as $V_A < \Vabar$, one obtains $S = -\log(|\mathcal{H}_A|)$, which is indeed the thermal entropy for subsystem $A$ at infinite temperature. Since random pure states mimic eigenstates at infinite temperature (i.e.\ $|\psi\rangle_{\beta = 0}$), this again suggests that the condition $V_A < \Vabar$ is perhaps sufficient, at least for some operators. 

On the other hand, there is a well-known Class I operator for which ETH fails when the ratio of subsystem to total system size $f = V_A/V$ is finite \cite{footnote:david}.  When $f$ is finite, the energy variance of the reduced density matrix $\rho_A(\ket{\psi}_\beta)$ will be suppressed by a factor of $(1-f)$ compared with the value the variance would have taken in the canonical ensemble.  Ultimately, this is due to the fact that a single eigenstate has precisely zero energy variance $\braket{(H - \braket{H})^2}$ in the full system, unlike the canonical ensemble, where the variance scales proportionally with system size.  This relationship can be expressed as
\be
\tr [ \rho_A(\ket{\psi}_\beta) \mathcal{O}_{A,\beta} ] = \frac{\Vabar}{V}\ \tr [ \rho_{A,\mathrm{th}}(\beta) \mathcal{O}_{A,\beta} ], \label{eq:suppressed_variance}
\ee
where $\mathcal{O}_{A,\beta} = (H_A - \braket{H_A}_\beta)^2$ is the energy variance operator.  We will explore implications of the subsystem energy variance mismatch more carefully in Sec.\ \ref{sec:results_finite_ratio}.

It is worth noting that for the case of time-evolved states, the full system variance is independent of time.  For a given initial state, this variance may indeed be different from the energy variance expected in the canonical ensemble, which implies that the energy variance for any subsystem that is a finite fraction of the total system will disagree, even at long times \cite{rigol2014}.  However, we expect that for ``typical'' initial states (which are typically inaccessible from an experimental point of view), the overall energy variance will match its result in the canonical ensemble, and so the energy variance for any subsystem will also match after thermalization (``canonical typicality'' \cite{tasaki98, goldstein2006, popescu2006}).

Overall, while we expect that ETH is obeyed by many Class I operators when $f > 0$ is finite, it cannot be satisfied by all such operators, since the subsystem energy variance provides an important counterexample.  Nonetheless, we expect that all Class I operators \emph{not} related to energy conservation (or another conserved quantity) should satisfy ETH as given in Eq.\ \ref{eq:eth1}. A more precise conjecture along these lines is that the set of operators spanned by $H^n_A$ where $n$ ranges between unity and the size of the Hilbert space do not satisfy ETH in the sense of Eq.\ \ref{eq:suppressed_variance} above. The number of such operators is still exponentially smaller than the total number of independent operators in a subsystem (e.g. in a spin-1/2 spin system, the total number of operators in a region $A$ is $4^{V_A}$, while the number of operators of the form $H^n_A$ is $2^{V_A}$, the size of the Hilbert space in region $A$).

\subsubsection{ETH for Class II operators} \label{sec:classII}

The extra ingredient introduced by Class II operators is that if ETH holds for them, then taking such an operator's expectation value with respect to a state $|\psi\rangle_\beta$ allows one to access the properties of the Hamiltonian at a temperature different than $\beta^{-1}$. For example, the Renyi entropy $S_\alpha$ corresponding to $\rho_A(|\psi\rangle_{\beta})$ satisfies $S_\alpha =  \frac{\alpha}{\alpha-1} V_A \beta \left( f(\alpha\beta) - f(\beta) \right)$, thus allowing one to access the free energy density at temperature $(\alpha \beta)^{-1}$.

Let us first consider the validity of ETH for Class II operators when $V_A \ll V$.  Remarkably, the results presented in the remainder of this paper demonstrate that ETH is valid for all Class II operators in this limit.  Thus, a single eigenstate of finite energy density contains knowledge of the properties of the system at all temperatures.

Now let us turn to the case in which $V_A/V$ is finite, which turns out to be much more subtle.  As mentioned in the previous subsection, there is already a Class I operator for which ETH fails in this limit, namely the subsystem energy variance.  Thus, we do not expect that ETH will hold for all Class II operators either when $f$ is finite.  In addition, for a given ratio $V_A/V$ with both $V_A,V \rightarrow \infty$, there is a physical constraint on the range of energy densities for which the spectrum of  $|\psi\rangle_\beta$ in principle can match that of $\rho_{A,\mathrm{th}}(\beta)$. To appreciate this, let us consider a slightly different problem---an arbitrary Hamiltonian of hardcore bosons with particle number conservation, at \textit{infinite temperature}. We will consider an explicit example of such a system in the next section. Since the total particle number operator $\hat{N}$ commutes with the Hamiltonian and satisfies the equation $\hat{N} = \hat{N}_A + \hat{N}_{\Abar}$, the reduced density matrix $\rho_A$ for a wavefunction $|\psi\rangle_{\beta = 0}$ is block diagonal in the number of particles $N_A$ in subsystem $A$. Furthermore, if ETH holds (as given by a generalization of Eqs.\ \ref{eq:eth2} and \ref{eq:eth3}), then the Schmidt decomposition is given by

\be 
|\psi\rangle_{\beta = 0} = \sum_{N_A=0}^{N} \sqrt{\lambda_{N_A}}\sum_{i}|u_{i}\rangle_{N_A} \otimes |v_{i}\rangle_{N-N_A} \label{eq:schmidtN}
\ee
where $\lambda_{N_A}$ are the Schmidt coefficients in the sector $N_A$, and $|u_{i}\rangle_{N_A}$, $|v_{i}\rangle_{N-N_A}$ are the corresponding eigenvectors. The label $i$ captures fluctuations of particles within a fixed sector $N_A$. Note that there is no index $i$ on $\lambda_{N_A}$ because we are at infinite temperature and all Schmidt states within a sector $N_A$ are equally likely. 

The decomposition in Eq.\ \ref{eq:schmidtN} allows one to calculate properties of subsystem $A$ at infinite temperature even \emph{away} from filling $N/V$ since the reduced density matrix $\rho_A$ will contain sectors with various densities $N_A/V_A$. However, there is both an upper limit and a lower limit on the density in subsystem $A$, since

\be 
 \textrm{max}\left[ N-(V-V_A),0 \right] \leq N_A \leq \textrm{min}\left[N,V_A\right] \label{eq:nalimits}
\ee
And thus the particle density $N_A/V_A$ in subsystem $A$ satisfies

\be 
\textrm{max}\left[ 1- (1-n)/f,0 \right] \leq \frac{N_A}{V_A} \leq \textrm{min}\left[n/f,1\right] \label{eq:densityrange}
\ee
where $n \equiv N/V$ is the overall particle density and $f \equiv V_A/V$. Thus, a necessary condition for the wavefunction in Eq.\ \ref{eq:schmidtN} to encode properties of the system at \textit{all} fillings is

\be 
f \leq \textrm{min}\left[ n, 1-n\right] \label{eq:nconstraint}
\ee

The above discussion, with some modifications, carries to systems with (only) energy conservation, at an arbitrary temperature. The Schmidt decomposition of an eigenstate $ |\psi\rangle_{\beta}$ with eigenvalue $E$  may now be written as:

\be 
|\psi\rangle_{\beta} = \sum_{i}  \sqrt{\lambda_i} |u_i\rangle\otimes |v_i\rangle  \label{eq:schmidt}
\ee
The physical content of ETH, as approximated in Eq.\ \ref{eq:eth3}, is that $\lambda_{i} \propto e^{-\beta E_{A,i}}$ where $E_{A,i}$ is the $i$'th energy eigenvalue of $H_A$ (the projection of the Hamiltonian to subsystem $A$) while $ |u_i\rangle$ is the corresponding eigenstate of $H_A$. Denoting the ground state energy to be zero, one naively expects that $\langle  u_i|H_A| u_i\rangle \leq E \,\, \forall \,\,| u_i\rangle$ since the energy density in the subsystem $\Abar$ cannot be less than the ground state energy density. However, this argument has a loophole since in contrast to the particle number operator $\hat{N}$, the total Hamiltonian is \textit{not} separable into subsystems $A$ and $\Abar$: $H = H_A + H_{\Abar} + H _{A \Abar} $, which actually allows $\langle  u_i|H_A| u_i\rangle$ to exceed $E$ as we will see in Sec.\ \ref{sec:results_finite_ratio} in the context of the model Hamiltonian in Eq.\ \ref{eq:tfi} below.  To understand the constraint on $\langle  u_i|H_A| u_i\rangle$ precisely, let us derive an expression which encapsulates the classical notion that the sum of energies in subsystem $A$ and $\Abar$ equals $E$.

We first note:

\bea
\langle u_{i0}| \otimes \langle v_{i0}|H|\psi\rangle_\beta & = & E\,\langle u_{i0}| \otimes \langle v_{i0}|\psi\rangle_\beta \\
& = & E \sqrt{\lambda_{i0}}
\eea

The above expression can  be re-evaluated  using the decomposition $H = H_A + H_{\Abar} + H _{A \Abar} $:
\bea 
& & \langle u_{i0}| \otimes \langle v_{i0}|H|\psi\rangle_\beta    \\ &= & \langle u_{i0}| \otimes \langle v_{i0}|H_A + H_{\Abar} + H_{A\Abar}|\psi\rangle_\beta \nonumber \\
& = & \sqrt{\lambda_{i0}} \langle u_{i0}|H_A|u_{i0}\rangle + \sqrt{\lambda_{i0}} \langle v_{i0}|H_{\Abar}|v_{i0}\rangle + \nonumber \\
&  & \sum_{j}  \sqrt{\lambda}_j \langle u_{i0}| \otimes \langle v_{i0}|H_{A\Abar}| u_j\rangle \otimes |v_j\rangle
\eea 

Equating the two ways to calculate the same expression, one finds:

\bea
& & \langle v_{i0}|H_{\Abar}|v_{i0}\rangle + \sum_{j}  \sqrt{\frac{\lambda_j}{\lambda_{i0}}} \langle u_{i0}| \otimes \langle v_{i0}|H_{A\Abar}| u_j\rangle \otimes |v_j\rangle \nonumber \\ 
& & = E - \langle u_{i0}|H_A|u_{i0}\rangle  \label{eq:energybalance}
\eea

Due to the variational principle for the ground state,  $\langle v_{i0}|H_{\Abar}|v_{i0}\rangle \geq -c L^{d-1}$ where $c$ is a constant (recall that in our convention, the ground state energy for the full Hamiltonian is set to zero). Since both $E$ and $\langle u_{i0}|H_A|u_{i0}\rangle$ scale as $L^d$, the only way for $\langle u_{i0}|H_A|u_{i0}\rangle$ to exceed $E$ is that the second term on the left hand side of Eq.\ \ref{eq:energybalance}, viz.\ $E_{\textrm{boundary}} \stackrel{\textrm{def}}{=} \sum_{j}  \sqrt{\frac{\lambda_j}{\lambda_{i0}}} \langle u_{i0}| \otimes \langle v_{i0}|H_{A\Abar}| u_j\rangle \otimes |v_j\rangle$, is negative and scales as $L^d$. When that happens,  ETH no longer holds, as we now argue on general grounds, and will also demonstrate numerically for a lattice Hamiltonian in Sec.\ \ref{sec:results_finite_ratio}. To see this, we reiterate that ETH requires that (i) $|u_i\rangle$'s are approximate eigenstates of $H_A$, and (ii)  $\lambda_{i} \propto e^{-\beta \langle u_i|H_A|u_i\rangle} = e^{-\beta E_{A,i}}$. Firstly, when $\langle u_{i0}|H_A|u_{i0}\rangle < E$ so that ETH could in principle hold, the  $E_{\textrm{boundary}}$ term can be neglected because the `diagonal term' in $E_{\textrm{boundary}}$ (i.e.\ the term corresponding to $j = i0$) scales as the boundary ($\propto L^{d-1}$) and is thus subleading, while the off diagonal terms scale as $e^{-L^d}$ and thus vanish in the thermodynamic limit (recall that $V_{\overline{A}} > V_A$).  On the other hand, when $\langle u_{i0}|H_A|u_{i0}\rangle > E$, the $|v_{i0}\rangle$'s now correspond to states of zero energy density, and the aforementioned argument for neglecting off-diagonal terms is no longer valid. So, let us assume that  $\langle u_{i0}|H_A|u_{i0}\rangle > E$ and each $|u_{i0}\rangle$ continues to be an eigenstate of $H_A$. Thus, one requires that 
\be 
\int\,de'\, \sqrt{\frac{\lambda(e')}{\lambda(e)}} M(e,e') e^{S(e')} \propto g(e)/L^{d-1}, \label{eq:energybalance2}
\ee
where we have taken the continuum limit and $\lambda(e)$ denotes the Schmidt eigenvalue corresponding to an eigenvector $|u\rangle$ at energy density $e$, while $M(e,e') = \langle u(e) | \otimes \langle v(e)|H_{A\overline{A}}|u(e') \rangle \otimes |v(e')\rangle$ and $g(e) = e - \langle u(e)|H_A|u(e)\rangle/L^d$. It is obvious from Eq.\ \ref{eq:energybalance2} that $\lambda(e) \propto  e^{-\beta E_{A}} = e^{-\beta e f L^d}$ is no longer the solution. In fact, the only way for the integral on the left hand side of Eq.\ \ref{eq:energybalance2} not to have any exponential dependence on $L$ (as required by the right hand side) is that the integrand itself does not have such dependence, i.e.\ $\sqrt{\frac{\lambda(e')}{\lambda(e)}} \propto \frac{1}{M(e,e')} e^{-S(e')}$. This implies a breakdown of ETH when $\langle u_{i0}|H_A|u_{i0}\rangle > E$. 

The above discussion implies that for a given wavefunction and bipartition, the maximum energy density that is potentially accessible in a subsystem $A$, such that the corresponding Schmidt weight  satisfies ETH is,
\be 
e^{*} = \textrm{min}(E/V_A, e_{\mathrm{max}}) = \textrm{min}(e/f,e_{\mathrm{max}}) \label{eq:ecrit}
\ee
where $e= E/V$ is the energy density corresponding to the wavefunction and  $e_\mathrm{max}$ is the maximum energy density for the Hamiltonian $H$ (recall that $e_\mathrm{max}$ can be finite for lattice-regularized quantum systems, e.g.\ for models of fermions or spins/hardcore bosons). Above, we have assumed that $ e< e_{\mathrm{max}}/2$. In the case when $e > e_{\mathrm{max}}/2$, the range of available energies is instead bounded from below by $\textrm{max}\left[0,e_{\mathrm{max}}(1- 1/f) - e/f \right]$. If our goal is to capture the fluctuations in the system for all energy densities so that \textit{all} Class II operators not related to  energy conservation satisfy ETH, we obtain an analog of Eq.\ \ref{eq:nconstraint} for the energy: $E/V_A \geq e_\mathrm{max}$, and, $(e_\mathrm{max}V - E)/V_A \geq e_\mathrm{max}$. Expressed in terms of the fraction $V_A/V$, and the energy density of the eigenstate $e = E/V$, this constraint is

\be 
f \leq f^{*} \equiv \textrm{min} \left[\frac{e}{e_\mathrm{max}}, 1- \frac{e}{e_\mathrm{max}}\right]. \label{eq:econstraint}
\ee

Let us emphasize that the above constraint is a \emph{necessary} condition for ETH to hold for all Class II operators, not a sufficient one.  Just as some Class I operators cannot satisfy ETH when $f$ is finite, we expect that there also exist Class II operators for which ETH fails when $f$ is finite, even when the above condition holds.  Even so, significant deviation in the eigenvalue spectrum begins where this constraint breaks down, as our numerical results will demonstrate in Sec.\ \ref{sec:results_finite_ratio}.

\begin{figure*}[htb]
    \centering
    \begin{tabular}{cc}
        \includegraphics[width=0.5\textwidth]{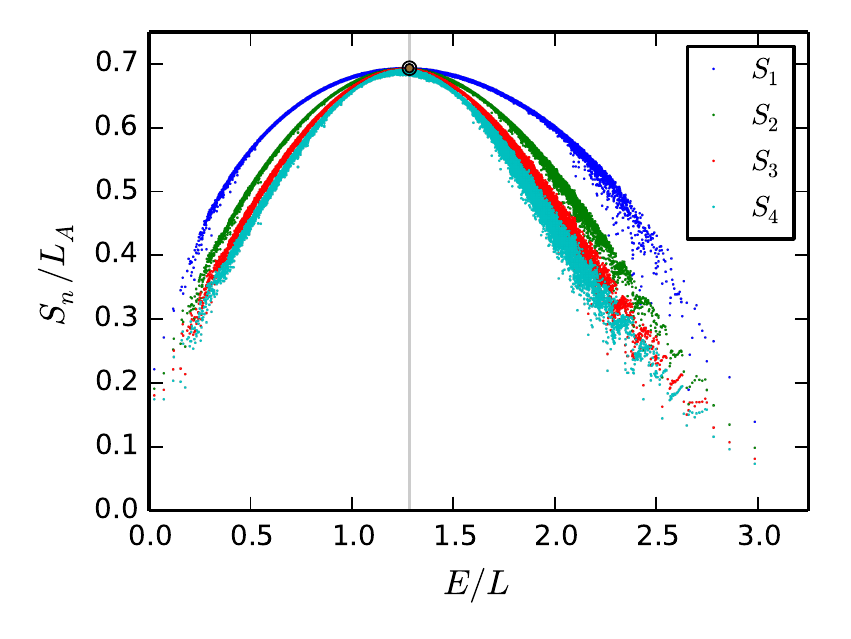} &
        \includegraphics[width=0.5\textwidth]{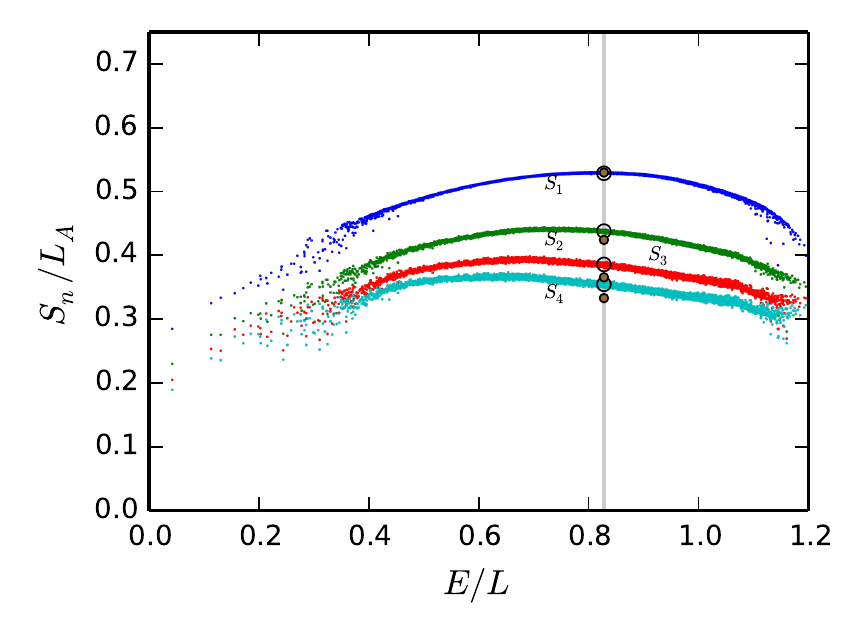}
    \end{tabular}
    \caption{Entanglement entropies $S_1$ through $S_4$ for a model with no conservation law (left panel, given by Eq.\ \ref{eq:tfi} at $L=21$), and a model with particle number conservation (right panel, given by Eq.\ \ref{eq:mu_model} at $L=27$ with filling $N=6$).  We use the parameters mentioned in the text to place each model at a nonintegrable point.  In each case we consider eigenstates in the $k=1$ sector, with subsystem size $L_A=4$.  The grey vertical line denotes infinite temperature (point of maximum $S_1$), and the black circles mark the theoretical predictions for the entanglement entropies there.  The brown markers denote the theoretical values of the entropies in the limit $L_A, L \rightarrow \infty$ while $L_A/L \rightarrow 0$, as given by Eqs.\ \ref{eq:mu_model_exact_S_alpha} and \ref{eq:mu_model_exact_S_1}.  Notice that the Renyi entropies all match at infinite temperature if and only if there are no additional conservation laws besides energy.}
    \label{fig:mu_model_entropy}
\end{figure*}

\subsection{Summary}  \label{sec:summarygeneral}

Let us summarize the discussion in this section.

\textbf{1. We conjecture  that ETH holds for all local and non-local Class I operators as long as $V_A \ll V$.}  This implies that ETH is \emph{not} restricted only to few-body operators (as can be seen in the limit $V_A/V \rightarrow 0$ as $V_A, V \rightarrow \infty$).  When the subsystem is taken to be a finite fraction $f < \frac12$ of the total system size, we provide some evidence in Sec.\ \ref{sec:results_finite_ratio} that all operators not involving energy conservation satisfy ETH as well.

 \textbf{2. We conjecture that ETH also holds for all Class II operators when $V_A \ll V$.}  It follows that a single eigenstate contains information about all energy densities available to the system.  When $f$ is finite, we suspect that ETH also holds for all operators that probe the system below an energy density $e^*$ (given by Eq.\ \ref{eq:ecrit}) and that do not involve energy conservation; however, we leave this as an open question.

 \textbf{3. Determining the full Hamiltonian from a single eigenstate is equivalent to the satisfaction of Eq.\ \ref{eq:eth1} for both Class I and Class II operators}.  Our results provide strong evidence that this is true when $V_A \ll V$.  Therefore, one should be able to extract information about the full Hamiltonian at arbitrary energy densities/temperatures using a single eigenstate.

\section{A warmup: eigenstates at infinite $T$}  \label{sec:mumodel}

\subsection{Von Neumann and Renyi entropy}

By definition, the thermal entropy reaches a maximum at infinite temperature.  Together with Eq.\ \ref{eq:SvN}, this implies that when ETH holds, eigenstates at ``infinite temperature'' are ones where the entanglement entropy is at its maximum.  Consider a 1D transverse field Ising model with longitudinal field, $H = \sum_{i=1}^{L} \left(\sigma_i^{z} \sigma_{i+1}^{z} + h_x \sigma_i^{x} + h_z \sigma_i^{z} \right)$.  Here the von Neumann entropy $S_1$ takes its maximum possible value when the eigenvalues of the reduced density matrix are all equivalent to one another.  Thus, from counting the basis size of the reduced Hilbert space, we expect for infinite temperature eigenstates that each eigenvalue of the reduced density matrix will approach $2^{-L_A}$ in the thermodynamic limit when $f = L_A / L < \frac12$.  From this, it follows that the Renyi entropies at infinite temperature satisfy
\be
S_\alpha = L_A \log 2,
\ee
that is, they are independent of Renyi index $\alpha$. The left panel of Fig.\ \ref{fig:mu_model_entropy} shows how the entropies $S_1$ through $S_4$ together match this predicted value at the infinite temperature point for a $L=21$ system with periodic boundary conditions and subsystem size $L_A=4$.  In general, as $L \rightarrow \infty$ the $T = \infty$ entropy density is given by $S_\alpha/L_A = \log 2$.

Now let us instead consider a model with an additional conservation law, namely particle number conservation.  Consider a 1D chain of hardcore bosons
\be
\begin{split}
    H = & -\sum_i \left( t b^\dagger_i b_{i+1} + t' b^\dagger_i b_{i+2} + \mathrm{H.c.} \right) \\ & + \sum_i \left( V n_i n_{i+1} + V' n_i n_{i+2} \right)
\end{split}
\label{eq:mu_model}
\ee
where $n_i \equiv b^\dagger_i b_i$.  We focus on this system with periodic boundary conditions at the non-integrable point $t=V=1$ and $t'=V'=0.96$.  This model was previously studied and shown to exhibit ETH in Refs.\ \onlinecite{santos2010,deutsch2013}.

\begin{figure}[b]
    \centering
    \includegraphics[width=0.5\textwidth]{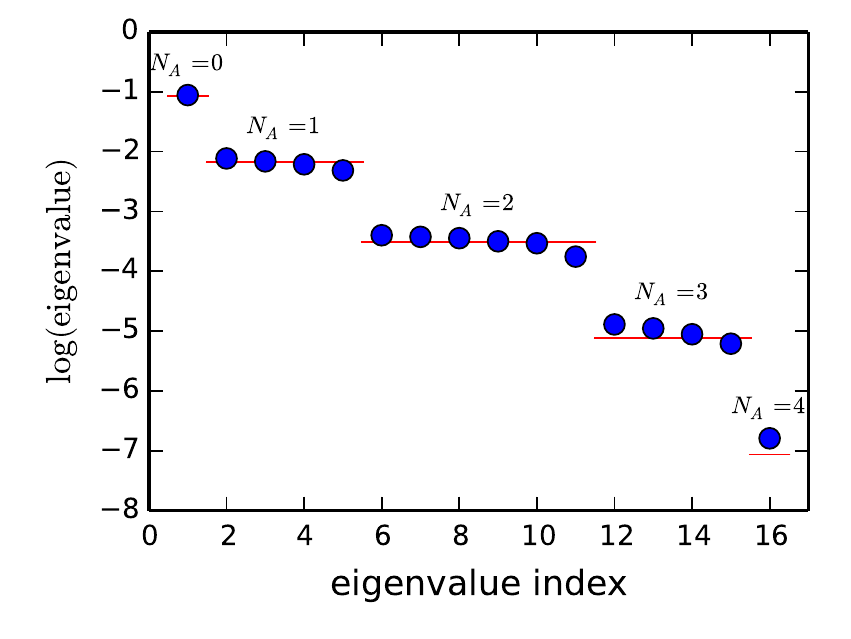}
    \caption{Eigenvalue spectrum of the reduced density matrix of an infinite temperature eigenstate, $\rho_A(\ket{\psi}_{\beta=0})$ for the hardcore boson model Eq.\ \ref{eq:mu_model} with $L=27$, $L_A=4$, and filling $N=6$.  The red lines plot the theoretical value of each eigenvalue in the thermodynamic limit, determined from the filling $N_A$ of the sector in which it lies.}
    \label{fig:mu_model_spectrum}
\end{figure}

Due to particle number conservation, the reduced density matrix from any pure state is block diagonal, with each block corresponding to some filling number $N_A$ of the subsystem $A$.  The block of the reduced density matrix $\rho_A^{(N_A)}$ corresponding to filling $N_A$ is a $d_{N_A} \times d_{N_A}$ matrix, where $d_{N_A} \equiv \binom{L_A}{N_A}$. At infinite temperature and for $L_A/L < \frac12$, the eigenvalues of $\rho_A$ must be equal to one another within a given block, but the eigenvalues in different blocks will be different: they are in fact proportional to $\binom{L-L_A}{N-N_A}$, the number of microstates consistent with such a configuration in subsystem $A$. Taking into account that $\tr(\rho_A) = 1$, one finds that each of the $d_{N_A} = \binom{L_A}{N_A}$ eigenvalues of $\rho_A^{(N_A)}$ are given by $\lambda_{N_A} \equiv \binom{L-L_A}{N-N_A}/\binom{L}{N}$. The spectrum of $\rho_A$ we find for a single eigenstate (as shown in Fig.\ \ref{fig:mu_model_spectrum}) is in agreement with that of the thermal reduced density matrix $\rho_{A,\mathrm{th}}(\beta=0)$ studied in Ref.\ \onlinecite{lauchli2013}, consistent with ETH.

With this, the von Neumann entropy at infinite temperature becomes
\be
S_1 = -\sum_{N_A} d_{N_A} \lambda_{N_A} \log \lambda_{N_A} 
\ee
and the Renyi entropies are given by
\be
S_\alpha = -\frac{1}{\alpha-1}\log\left( \sum_{N_A} d_{N_A} \lambda_{N_A}^\alpha \right), \label{eq:Snmumodel}
\ee
where the sums over $N_A$ are restricted to subsystem particle fillings $N_A$ that satisfy the constraint in Eq.\ \ref{eq:nalimits}.  The above expressions are valid when $L_A / L < \frac12$.

Because the eigenvalues are non-uniform, the Renyi entropies $S_\alpha$ at infinite temperature depend on the Renyi index $\alpha$, in contrast to an energy-only conserving model.  The right panel of Fig.\ \ref{fig:mu_model_entropy} shows how the actual values of $S_1$ through $S_4$ match those predicted by the above counting argument.

For comparison, we also calculate $S_\alpha$ analytically in the thermodynamic limit. For simplicity, we consider the limits, $L, N,L_A \rightarrow \infty$ such that $n = N/L$ is held constant, while $L_A/L \rightarrow 0$. In these limits, one can evaluate the expressions in Eq.\ \ref{eq:Snmumodel} using Stirling's approximation $\log(x!) \approx x \log(x) - x$. One finds that in the limits considered, $S_\alpha$ receives contribution only from $N_A$ given by

\be
N^{*}_A = \frac{L_A}{1+ \left(\frac{1}{n}-1\right)^\alpha}
\label{eq:mu_model_saddle_point}
\ee
Thus, $S_\alpha$ probes the system at filling $N_A^{*}/L_A = \frac{1}{1+ \left(\frac{1}{n}-1\right)^\alpha}$, which is different than the actual filling $n$, unless $\alpha=1$ (which corresponds to the von Neumann entanglement entropy). This also immediately leads to expressions for Renyi and von Neumann entanglement entropies in the thermodynamic limit:

\be 
S_\alpha / L_A = -\frac{1}{\alpha-1}\log\left[n^\alpha + (1-n)^\alpha\right]
\label{eq:mu_model_exact_S_alpha}
\ee

and 

\be 
S_1 / L_A = -\left[ n \log(n) + (1-n)\log(1-n)\right].
\label{eq:mu_model_exact_S_1}
\ee

We plot these values in Fig.\ \ref{fig:mu_model_entropy} for comparison. Remarkably, even with the small system sizes we can access, the difference between the exact finite size result (obtained by counting over all sectors) and the result valid in the thermodynamic limit is quite small.

In the above derivation, it is also possible to relax the restriction $L_A/L \rightarrow 0$ as $L_A, L \rightarrow \infty$.  We then find that $N_A^{*}$ is given by the solution to
\be
N_A^{*} = \frac{L_A}{1 + \left(\frac{1-f}{n - f N_A^{*} / L_A} - 1\right)^\alpha},
\ee
which reduces to Eq.\ \ref{eq:mu_model_saddle_point} when $f \rightarrow 0$.  

Let us note a few things about this equation:

\begin{enumerate}

\item When $\alpha=1$, the solution is $N_A^{*} = n L_A$, regardless of $f$.  Thus, the von Neumann entropy always probes the system at its given filling, even when $f$ is finite.  Further analysis shows that Eq.\ \ref{eq:mu_model_exact_S_1} holds generally when $f < \frac12$.

\item When the system is at half filling ($n=\frac12$), the solution is $N_A^{*} = \frac12 L_A$, regardless of $f$ or $\alpha$. 

\item When $\alpha > 1$, $0 < f < \frac12$, and $n \ne \frac12$, the filling fraction $N_A^{*}/L_A$ probed by the Renyi entropy $S_\alpha$ actually depends on $f$.  As a result, the Renyi entropies for a given $L_A$ depend on $f$.  This can be contrasted with the von Neumann entropy, which is independent of $f$ as long as $f < \frac12$.  The right panel of Fig.\ \ref{fig:mu_model_entropy} illustrates this nicely: the analytical $f \rightarrow 0$ prediction for the von Neumann entropy (Eq.\ \ref{eq:mu_model_exact_S_1}) matches the corresponding numerical result  quite well, but the Renyi entropies differ significantly because $f=4/27$ is finite.

\end{enumerate}
We expect that analogous features hold true also for the model that conserves only energy, which we will discuss in the later sections.

\subsection{Subsystem energy variance}

Let us also consider the average subsystem filling variance of the particle-number conserving system given by Eq.\ \ref{eq:mu_model} at infinite temperature.  While the average subsystem filling is given by $\braket{N_A} = n L_A = N f$, the variance in this quantity for a single eigenstate with $f \equiv L_A/L < \frac12$ is given by
\be
\Braket{(N_A - \braket{N_A})^2} = L_A (1-f) (1-n) \frac{L}{L-1}.
\ee
Although both the filling and its variance are proportional to $L_A$ as expected, the variance includes an additional factor $(1-f)$, which causes it to be suppressed compared with the grand canonical ensemble when $f$ is finite.  In Sec.\ \ref{sec:results_finite_ratio} we will witness a similar suppression of the subsystem energy variance when the condition $L_A/L \rightarrow 0$ is relaxed.

\begin{figure}[b]
    \centering
    \includegraphics[width=0.5\textwidth]{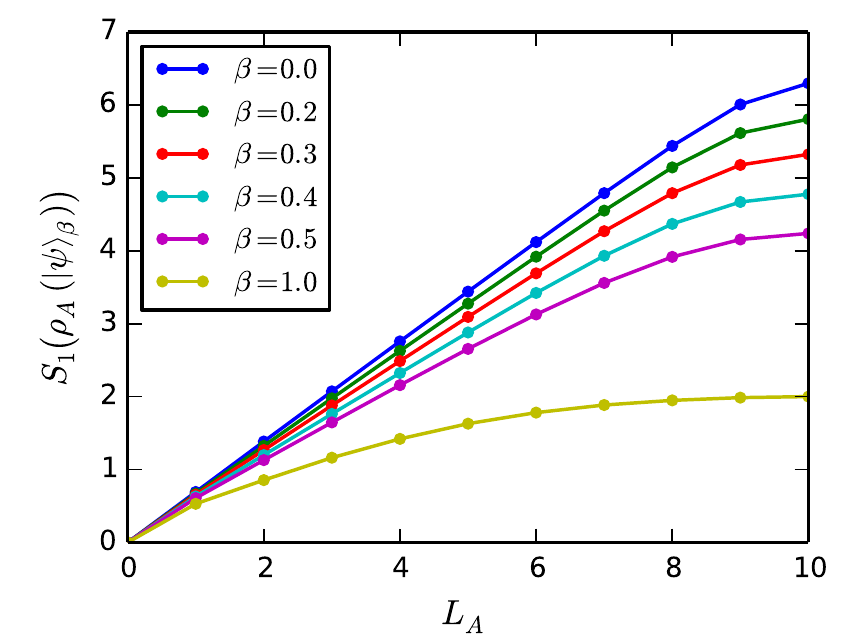}
    \caption{Scaling of the von Neumann entanglement entropy $S_1$ with subsystem size for the $L=20$ system given in Eq.\ \ref{eq:tfi}.  Up to $\beta=0.5$, the scaling is linear for small $L_A$, suggesting that the states are volume-law and are thus likely to satisfy ETH.  The $\beta=1.0$ eigenstate, on the other hand, is clearly not linear, and is too close to the ground state at this system size to exhibit ETH.}
    \label{fig:entropy_scaling}
\end{figure}

\begin{figure*}[ht]
    \centering
    \begin{tabular}{cc}
        \includegraphics[width=0.5\textwidth]{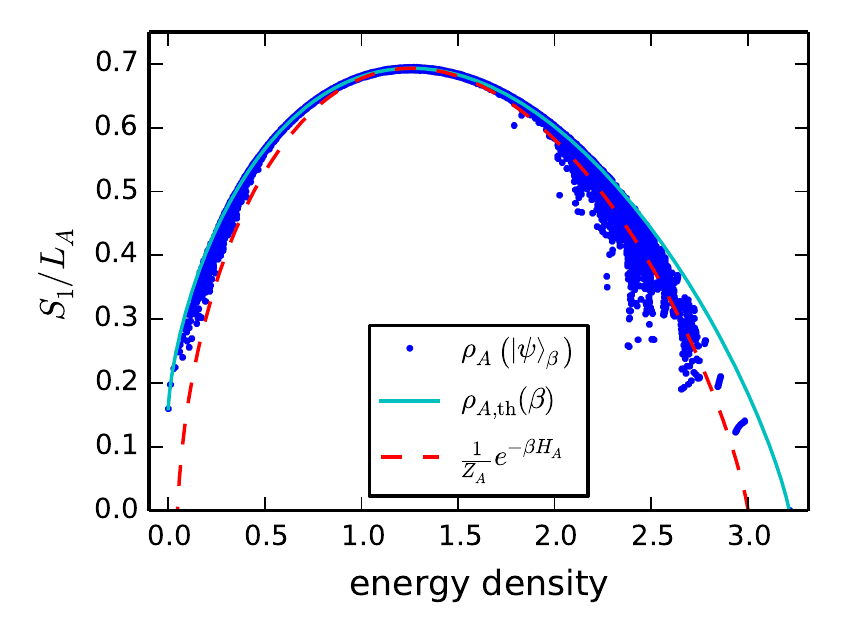} &
        \includegraphics[width=0.5\textwidth]{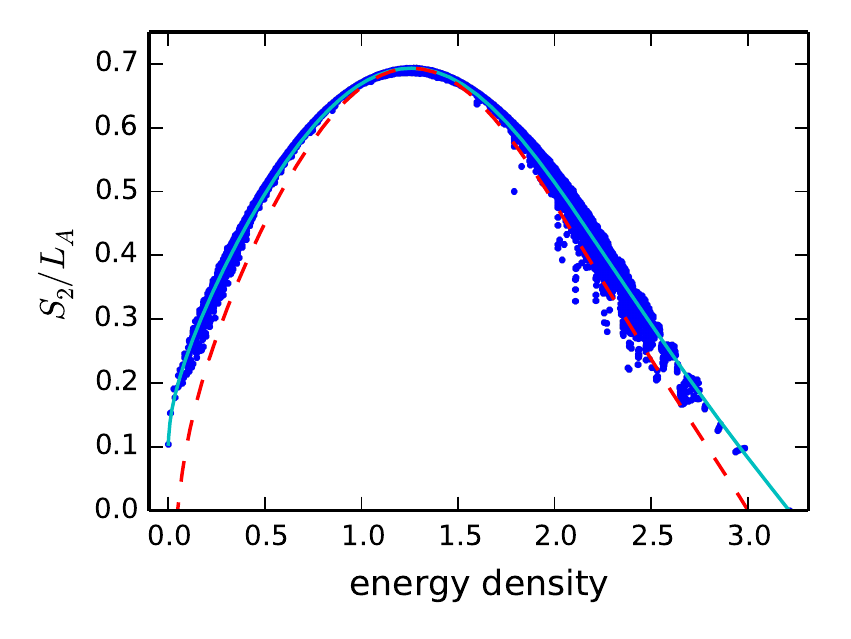} \\
        \includegraphics[width=0.5\textwidth]{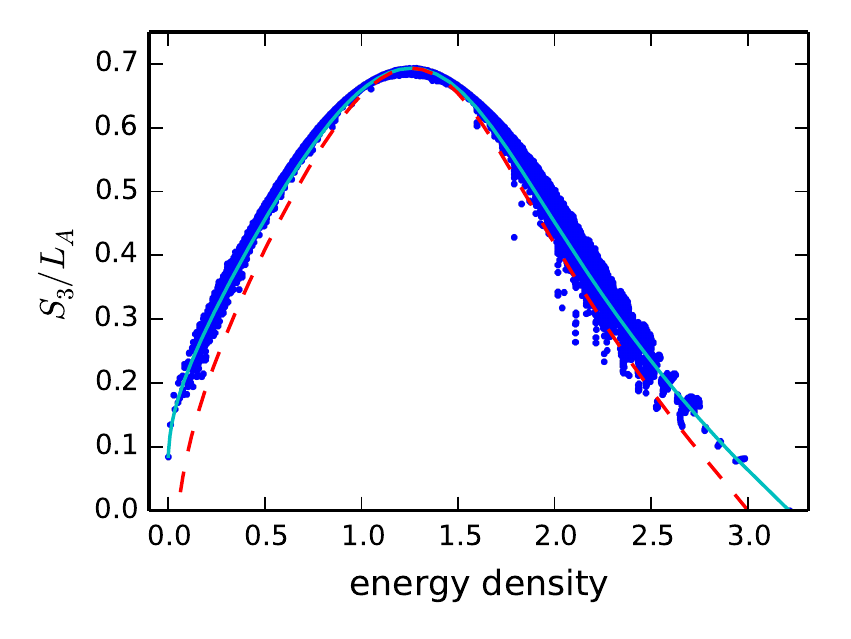} &
        \includegraphics[width=0.5\textwidth]{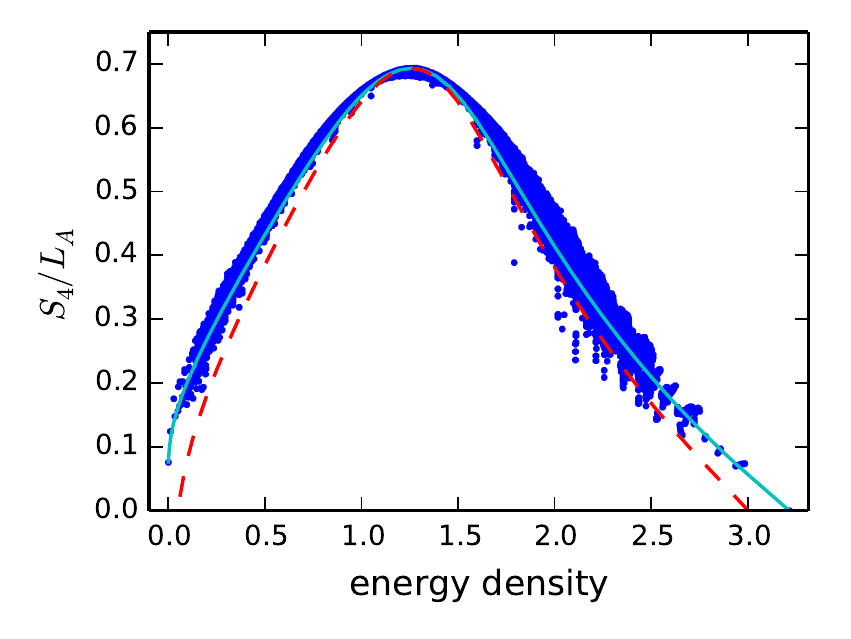} \\
    \end{tabular}
    \caption{The von Neumann entropy $S_1$ and Renyi entropies $S_2$, $S_3$, and $S_4$ for the system given in Eq.\ \ref{eq:tfi} with $L=21$ and $L_A=4$.  Here, $Z_A = \tr_A(e^{-\beta H_A})$.  The entropies of the reduced density matrix at each energy density agree remarkably with the the entropies calculated from the canonical ensemble, given by Eqs.\ \ref{eq:eth2} and \ref{eq:eth3}.}
    \label{fig:entropy}
\end{figure*}

\section{Model Hamiltonian with only energy conservation}   \label{sec:tfi}

To develop some understanding of the questions posed in the introduction, we study a finite 1D quantum spin-1/2 chain with the following Hamiltonian:

\be 
H = \sum_{i=1}^{L} \left(\sigma_i^{z} \sigma_{i+1}^{z} + h_x \sigma_i^{x} + h_z \sigma_i^{z} \right) \label{eq:tfi}
\ee
We set $h_x = 0.9045$ and $h_z = 0.8090$ such that the model is far away from any integrable point, and is expected to satisfy ETH in the sense of Eq.\ \ref{eq:eth1} as shown in Ref.\ \onlinecite{kim2014}.  We use periodic boundary conditions throughout.

We diagonalized the Hamiltonian in Eq.\ \ref{eq:tfi} for system sizes up to $L=21$, obtaining all eigenvalues and eigenstates. As hinted earlier, to each eigenstate we assigned a temperature $\beta^{-1}$ by finding the value $\beta$ for which the energy expectation value in the canonical ensemble matches the energy of the eigenstate:
\begin{equation}
\frac{\langle\psi|H|\psi\rangle}{\langle\psi\vert\psi\rangle} = \frac{\tr\left( H e^{-\beta H} \right)}{\tr\left( e^{-\beta H} \right)}.
\label{eq:beta_calculate}
\end{equation}
By definition, $\beta=+\infty$ for the ground state and $\beta=-\infty$ for the highest excited state.  In practice, the range of available $\beta$ values on a finite size system is much smaller.  With $L=21$, for instance, the first excited state has $\beta \approx 4.0$, and the second-to-highest excited state has $\beta \approx -0.6$ (as determined from Eq.\ \ref{eq:beta_calculate}).  It follows that eigenstates outside the range $4.0 \gtrsim \beta \gtrsim -0.6$ will not appear fully thermal due to the large thermal correlation length expected at low temperatures. (This can be seen for instance in Fig.\ \ref{fig:entropy_scaling}, where the finite size corrections to the linear scaling of the entanglement entropy become more prominent as temperature decreases.)  Another thing to consider is that the infinite temperature eigenstate $\ket{\psi}_{\beta=0}$ is completely random and contains no information about the Hamiltonian.  In a finite size system, states near infinite temperature will also contain little information about the Hamiltonian and will therefore be unable to predict properties of the system at other energy densities.  As a result of these finite size considerations, we typically study values of $\beta$ between $0.2$ and $0.5$ in the remainder of this paper.

\begin{figure}[tb]
    \centering
    \begin{tabular}{c}
    \includegraphics[width=0.5\textwidth]{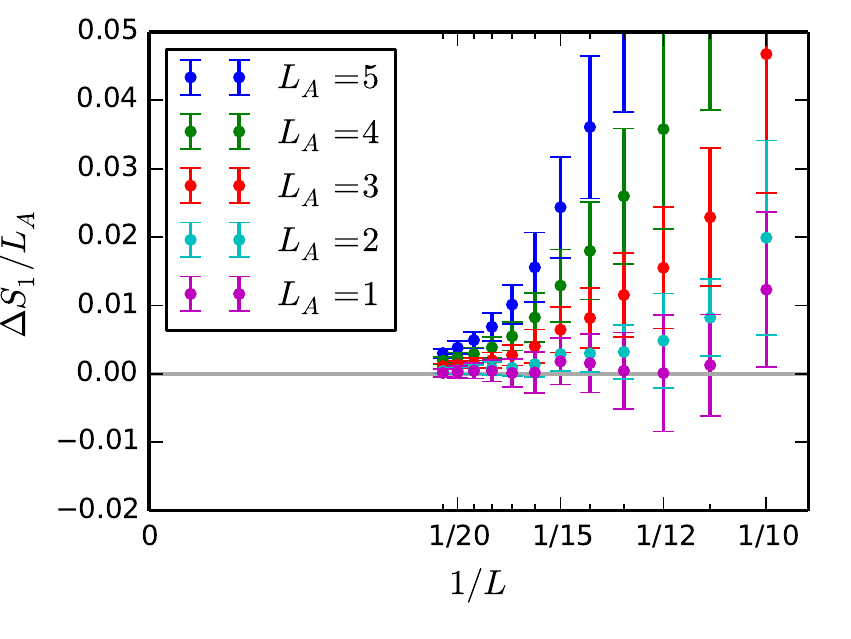} \\
    \includegraphics[width=0.5\textwidth]{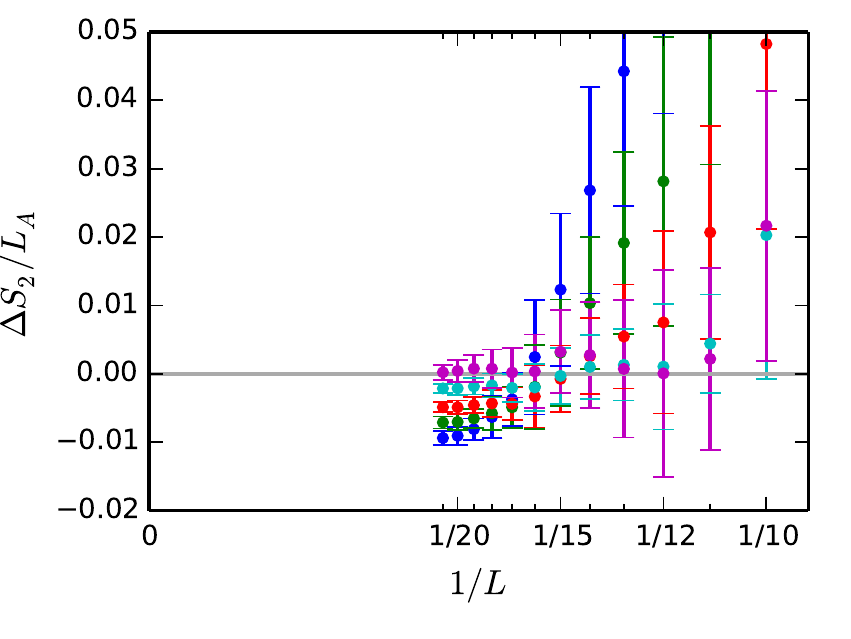}
  \end{tabular}
    \caption{Scaling of the entropy deviation $\Delta S_\alpha \equiv S_\alpha(\rho_{A,\mathrm{th}}(\beta)) - S_\alpha(\rho_A(\ket{\psi}_\beta))$ with $1/L$ for constant $L_A$ averaged over all eigenstates in the range $0.28 < \beta < 0.32$, for $S_1$ (top panel) and $S_2$ (bottom panel). The error bars represent one standard deviation away from the mean.  For $S_1$ this deviation is strictly non-negative, but for higher Renyi entropies it can oscillate and become negative before tending to zero as $L \rightarrow \infty$.}
    \label{fig:entropy_deviation_const_La}
\end{figure}

\section{Von Neumann and Renyi entropy of eigenstates at finite $T$}  \label{sec:Sn}

\subsection{ETH prediction for von Neumann and Renyi entropies}  \label{sec:ethSn}
Let us consider the Renyi Entropy $S_\alpha = -\frac{1}{\alpha-1}\log(\tr \,\rho^\alpha_A(|\psi\rangle_{\beta})) $ corresponding to an eigenstate $|\psi\rangle_{\beta}$ at inverse temperature $\beta$. Assuming that ETH, as encoded in Eq.\ \ref{eq:eth2}, holds, $S_\alpha$ may be reexpressed as:

\be 
S_\alpha = -\frac{1}{\alpha-1}\log\left(\frac{Z(A,\alpha,\beta)}{Z(1,\beta)^\alpha}\right)
\ee
where $Z(A,\alpha,\beta)$ is the partition function of the system on an $\alpha$-sheeted Riemann surface, such that subsystem $A$ has an effective temperature $(\alpha\beta)^{-1}$ while subsystem $\overline{A}$ has an effective temperature $\beta^{-1}$. $Z(1,\beta)$ is the regular partition function of the system \cite{holzhey1994, callan1994, calabrese2004}. Therefore, keeping terms \textit{only to the leading order in the subsystem size}, the above expression leads to Eq.\ \ref{eq:Sn} advertised in the Introduction,

\begin{figure}[tb]
    \centering
    \includegraphics[width=0.5\textwidth]{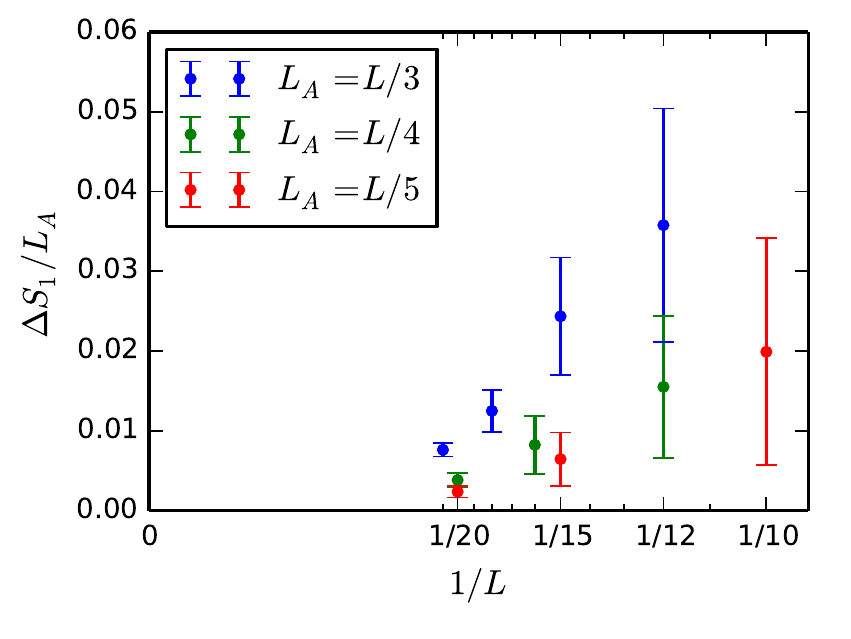}
    \caption{Scaling of the von Neumann entropy deviation $\Delta S_1$ with $1/L$ for constant ratio $L_A/L$ averaged over all eigenstates in the range $0.28 < \beta < 0.32$.  As in Fig.\ \ref{fig:entropy_deviation_const_La}, the error bars represent one standard deviation away from the mean.  Even though this plot considers the case where the subsystem size $L_A$ becomes infinite as $L \rightarrow \infty$, the entropy deviations are going to zero rapidly as $L$ becomes larger.}
    \label{fig:entropy_deviation_const_ratio}
\end{figure}

\begin{figure}[tb]
    \centering
    \includegraphics[width=0.5\textwidth]{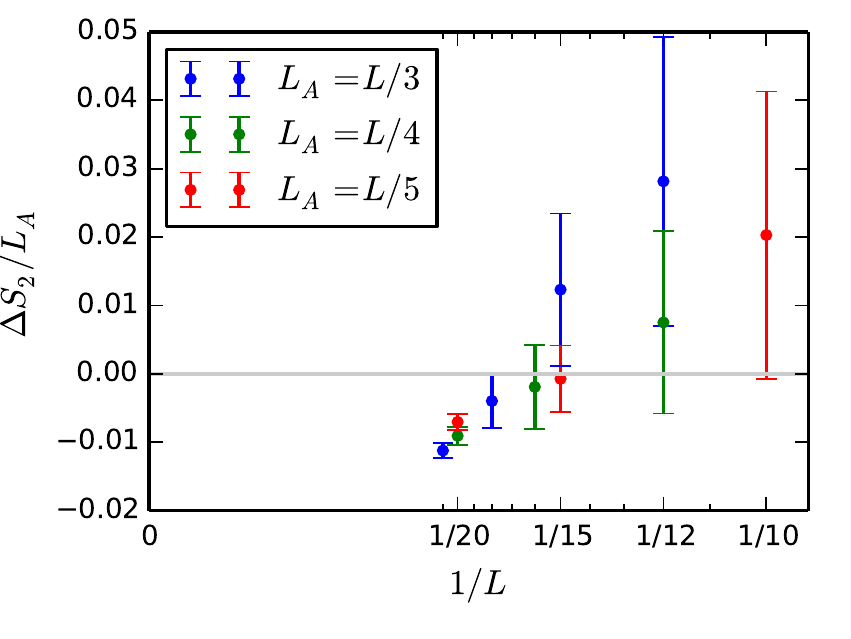}
    \caption{Scaling of the Renyi entropy deviation $\Delta S_2$ with $1/L$ for constant ratio $L_A/L$ averaged over all eigenstates in the range $0.28 < \beta < 0.32$.  As in Fig.\ \ref{fig:entropy_deviation_const_La}, the error bars represent one standard deviation away from the mean.}
    \label{fig:entropy_deviation_const_ratio_S2}
\end{figure}

\bea 
S_\alpha & = & -\frac{1}{\alpha-1} \log\left(\frac{e^{-\alpha \beta V_A f(\alpha\beta)-\alpha \beta V_{\overline{A}} f(\beta)}}{e^{-\alpha \beta V_A f(\beta)-\alpha \beta V_{\overline{A}} f(\beta)}}\right) \\
& = & \frac{\alpha}{\alpha-1} V_A \beta \left( f(\alpha\beta) - f(\beta) \right)
\eea

\noindent where $f$ is the free energy density. Therefore, the wavefunction at temperature $\beta^{-1}$ can be used to calculate the free energy at temperature $(\alpha\beta)^{-1}$. Indeed, the same result also follows using the approximate form in Eq.\ \ref{eq:eth3}. Taking the limit $\alpha\rightarrow 1$ leads to the conclusion that the von Neumann entanglement entropy $S_1$ satisfies 

\be 
S_1 = V_A s_\mathrm{th}(\beta) \label{eq:SvN}
\ee
where $s_\mathrm{th}(\beta) = S_1( \rho_{A,\mathrm{th}}(\beta) ) / L_A$ is the thermal entropy density at temperature $\beta^{-1}$.

\subsection{Numerical Results for von Neumann and Renyi entropies}  \label{sec:Snnumerical}
Fig.\ \ref{fig:entropy_scaling} shows the scaling of von Neumann entropy $S_1$ as a function of subsystem size $L_A$ for the eigenstates $|\psi\rangle_\beta$ of our model (Eq.\ \ref{eq:tfi}). As discussed in Sec.\ \ref{sec:classI}, we expect Eq.\ \ref{eq:SvN} to hold as long as $V_A < V_{\overline{A}}$, in the limit $V_A, \Vabar \rightarrow \infty$. This implies that in the thermodynamic limit, the function $S_1(V_A)$ is expected to form an inverted triangle shape, similar to the behavior of a random pure state (Eq.\ \ref{eq:EErand}). However, in a finite total system at any non-infinite temperature, $S_1$ is an analytic function of the ratio $V_A/V$ with a negative sign for $ \frac{d^2S_1}{dV_A^2}$, as shown in Fig.\ \ref{fig:entropy_scaling} (note that the sign of the curvature is fixed by the strong subadditivity of entanglement). However, even in finite system, the volume law does hold to a good accuracy when $V_A \lesssim V/2$, and the finite size scaling, discussed below, indicates that the inverted triangle shape is recovered in the thermodynamic limit.

Fig.\ \ref{fig:entropy} shows the comparison of $S_1$, $S_2$, $S_3$, and $S_4$ calculated for each individual eigenstate for a subsystem size $L_A = 4$ in a $L=21$ system, with their ETH predicted canonical counterparts, Eqs.\ \ref{eq:SvN} and \ref{eq:Sn}. We use two different canonical counterparts corresponding to Eqs.\ \ref{eq:eth2} and \ref{eq:eth3}, the latter version being susceptible to boundary errors, which nevertheless are expected to vanish as $V_A, V_{\overline{A}} \rightarrow \infty$.  The agreement for each entropy is remarkable. It is worth re-iterating that the Renyi entropies for an eigenstate $|\psi\rangle_\beta$ encode the free energy densities at temperatures different than $\beta^{-1}$ (Eq.\ \ref{eq:Sn}), and these results provide an instance of non-local Class II operators satisfying ETH. Also note that as $\alpha$ becomes larger, finite size effects become more pronounced because $S_\alpha$ probes the system at lower temperatures $(\alpha\beta)^{-1}$.

We also studied finite-size scaling of the von Neumann entropy and Renyi entropies by keeping $L_A$ constant and varying the total system size. The top panel of Fig.\ \ref{fig:entropy_deviation_const_La} shows the deviation $\frac{\Delta S_1}{L_A} = \frac{S_1(|\psi\rangle_\beta)}{L_A} - s_\mathrm{th}(\beta)$ for eigenstates in a range of temperatures. The difference $\Delta S_1/L_A$ seemingly goes to zero faster than any inverse power of $L$, and is consistent with an exponential dependence $\Delta S_1/L_A \sim e^{-L}$, or at the very least, a power-law decay $\Delta S_1/L_A \sim 1/L^{x}$ with $x \gg 1$ (although we should caution that inferring the precise asymptotic finite size scaling behavior using exact diagonalization studies is an inherently difficult task). The bottom panel shows a similar plot for the deviation of Renyi entropy $S_2$ from its ETH predicted value, Eq.\ \ref{eq:Sn}. The finite size scaling of $\Delta S_2$ is relatively difficult because unlike $S_1$, $S_2$ shows oscillations as a function of $L_A$ (see e.g. \cite{lauchli2013, calabrese2010}). Despite this, $\Delta S_2$ is less than a few percent of $S_2$ itself.

Fig.\ \ref{fig:entropy_deviation_const_ratio} plots the entropy deviation $\Delta S_1 / L_A$ for constant ratio $L_A/L$ at all available system sizes.  Although it is difficult to do a detailed scaling analysis with so few points, the data strongly suggests that $\Delta S_1 / L_A$ vanishes in the thermodynamic limit.

The finite size scaling of Renyi entropies at constant ratio $L_A/L$ is less conclusive, as can be seen in Fig.\ \ref{fig:entropy_deviation_const_ratio_S2}. The analytical argument for the particle number conserving model suggests that the Renyi entropies $S_\alpha$ for $\alpha \neq 1$ do not match their canonical counterparts when $V_A/V$ is held fixed. Ref.\ \onlinecite{liu2015} arrived at similar conclusions using a different approach.

\begin{figure}[b]
    \centering
    \includegraphics[width=0.5\textwidth]{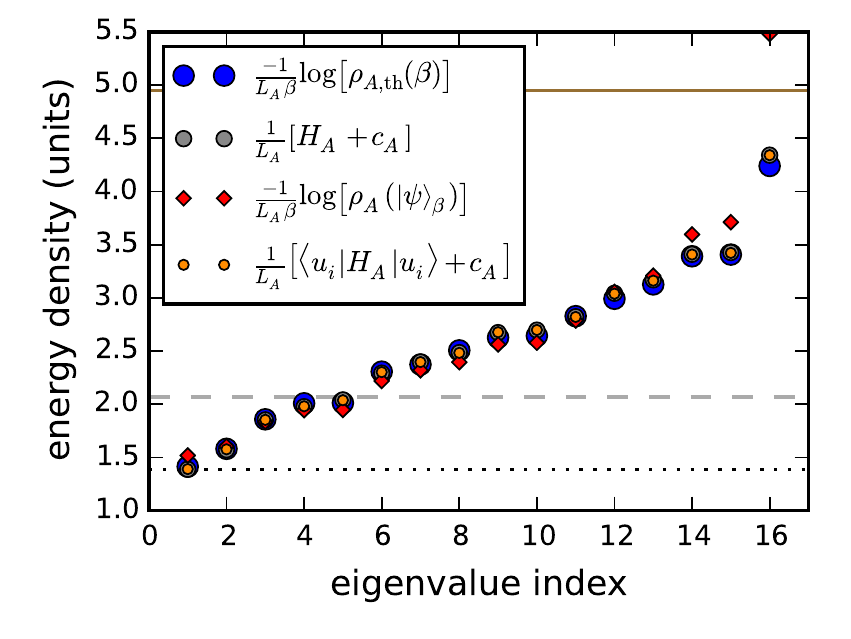}
    \caption{Comparison of the four quantities defined in the inset for an $L_A=4$ subsystem at $L=21$ and $\beta=0.3$.  Each quantity has been normalized so that the $y$-axis has units of energy density. The blue markers show the spectrum of the canonical (i.e.\ thermal) reduced density matrix while the red diamond markers correspond to the eigenvalues of a reduced density matrix $\rho_A(\ket{\psi}_\beta)$ for a \textit{single} eigenstate at temperature $\beta$; the grey markers show the eigenvalues of $H_A$ with a shift $c_A \equiv \frac{1}{\beta} \log Z_A = \frac{1}{\beta} \log \tr_A(e^{-\beta H_A})$ so that it can be directly compared with $-\frac{1}{\beta} \log[\rho_A(\ket{\psi}_\beta)]$ in accordance with Eq.\ \ref{eq:eth3} (note also that the combination $H_A + c_A$ is independent of the shift of the spectrum of $H_A$ by an arbitrary uniform constant). Finally, the orange markers represent the expectation value of $H_A$, again with a shift $c_A$, with respect to the Schmidt eigenvector $\left| u_i \right\rangle$ of $\rho_A(\ket{\psi}_\beta)$.  In each case, the eigenvalues/eigenvectors are ordered from smallest to largest energy density.  The horizontal lines plot the energy density $e$ (dashed, grey) and the critical energy density $e^* = \frac{eL}{L_A}$ (solid, brown) of the original eigenstate $\ket{\psi}_\beta$, with respect to the ground state energy density of $H_A + c_A$ (dotted, black).}
    \label{fig:spectra_overlap}
\end{figure}

\begin{figure*}[htb]
    \centering
    \begin{tabular}{cc}
        \includegraphics[width=0.5\textwidth]{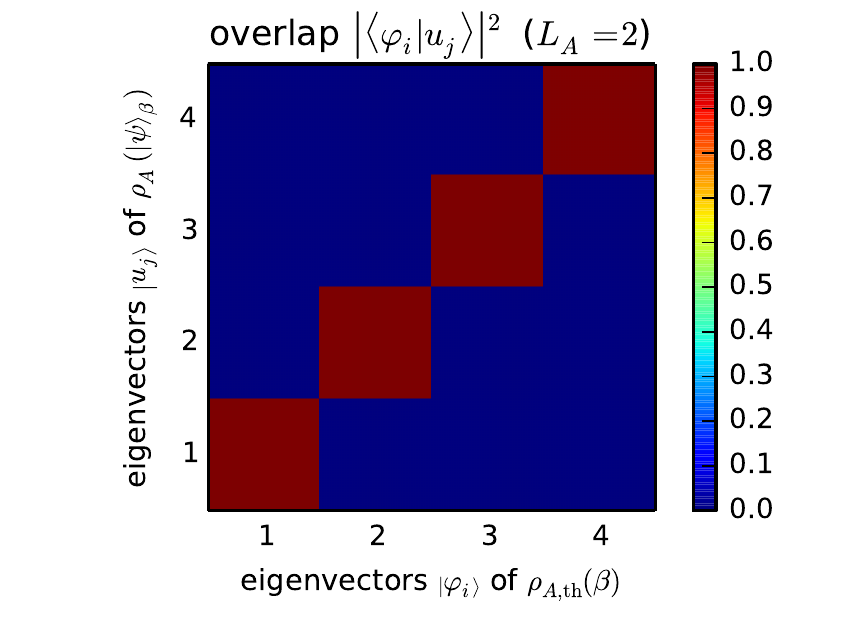} &
        \includegraphics[width=0.5\textwidth]{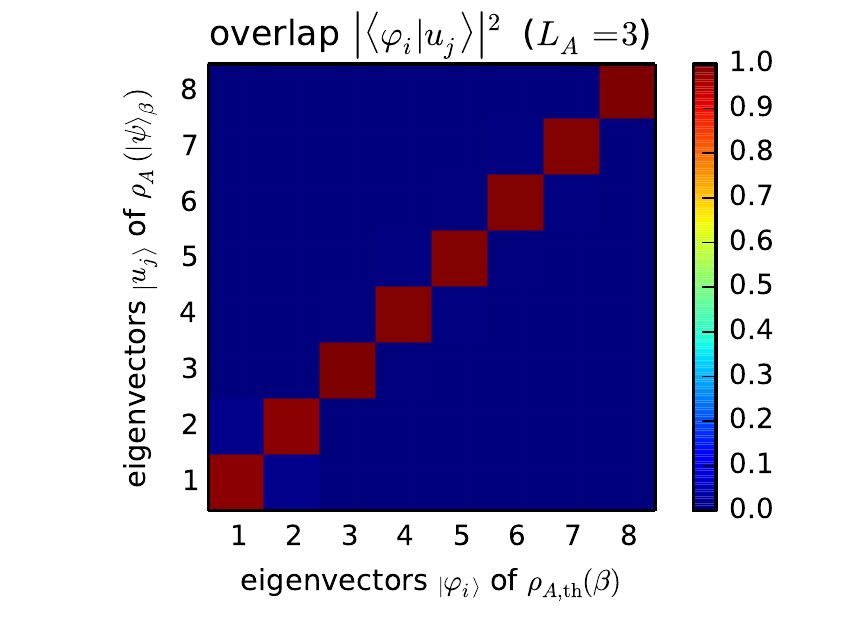} \\
        \includegraphics[width=0.5\textwidth]{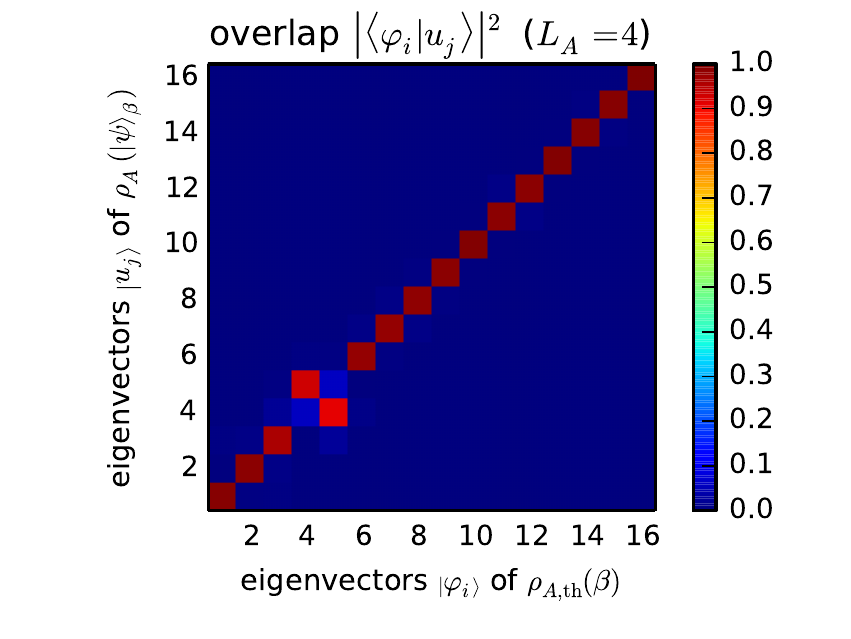} &
        \includegraphics[width=0.5\textwidth]{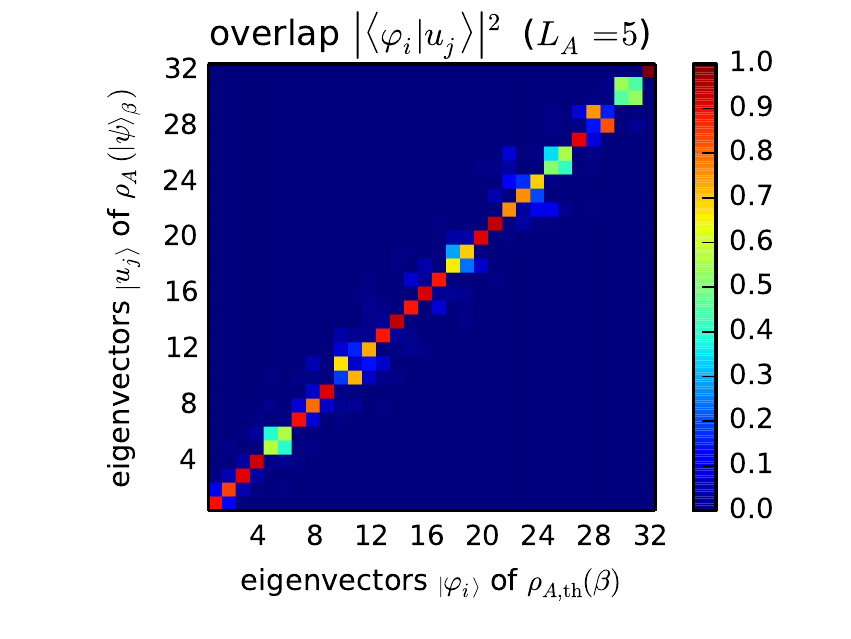} \\
    \end{tabular}
    \caption{Overlap between the Schmidt eigenvectors $\ket{u_j}$ and the eigenvectors $\ket{\varphi_i}$ of the canonical density matrix, for an $L=21$ system with $\beta=0.3$, and subsystem sizes $L_A=2,3,4,5$.  In each case, the eigenvectors are ordered from most significant (largest eigenvalue) to least significant (smallest eigenvalue).}
    \label{fig:eigenvector_overlap}
\end{figure*}

\section{Extracting the Hamiltonian from a single eigenstate} \label{sec:extracting}

In this section we will present numerical results that substantiate our conjecture that ETH is valid for all Class I and Class II operators when $V_A \ll V$ as $V \rightarrow \infty$.  Our numerical results consider the case where $V_A$ is held constant as $V \rightarrow \infty$.  We expect that all results in this section also hold true when the limits are taken such that $f \equiv V_A/V \rightarrow 0$ as $V_A, V \rightarrow \infty$.  In Sec.\ \ref{sec:results_finite_ratio} we will explore more carefully the case when $f < \frac12$ is finite.

We begin by probing in detail the entanglement spectra of individual eigenstates as well as the corresponding Schmidt states.  Specifically, we compare four different quantities, as shown in Fig.\ \ref{fig:spectra_overlap}, which test the validity of Eqs.\ \ref{eq:eth2} and \ref{eq:eth3}. The agreement of the spectrum of $\frac{-1}{\beta}\log[\rho_A(|\psi\rangle_{\beta})]$ with that of $\frac{-1}{\beta}\log[\rho_{A,\mathrm{th}}(\beta)]$ as well as with the actual Hamiltonian $H_A$ in region $A$ implies that essentially, the Schmidt eigenvalues $\lambda_i$ satisfy $\lambda_i \propto e^{-\beta E_{A,i}}$ where $E_{A,i}$ are the eigenvalues of $H_A$. Similarly, the agreement with the expectation value $\langle  u_i|H_A| u_i\rangle$ shows that the Schmidt eigenvectors $|u_i\rangle$ have the same character as the eigenvectors of the thermal density matrix.
 
To probe the Schmidt eigenvectors further, we directly calculated the overlaps between the eigenvectors of the reduced density matrix $\rho_A(|\psi\rangle_\beta)$ and the eigenvectors of the thermal density matrix $\rho_{A,\mathrm{th}}(\beta)$ (see Fig.\ \ref{fig:eigenvector_overlap}). Again, we find excellent agreement.

\begin{figure*}[tb]
    \centering
    \includegraphics[width=\textwidth]{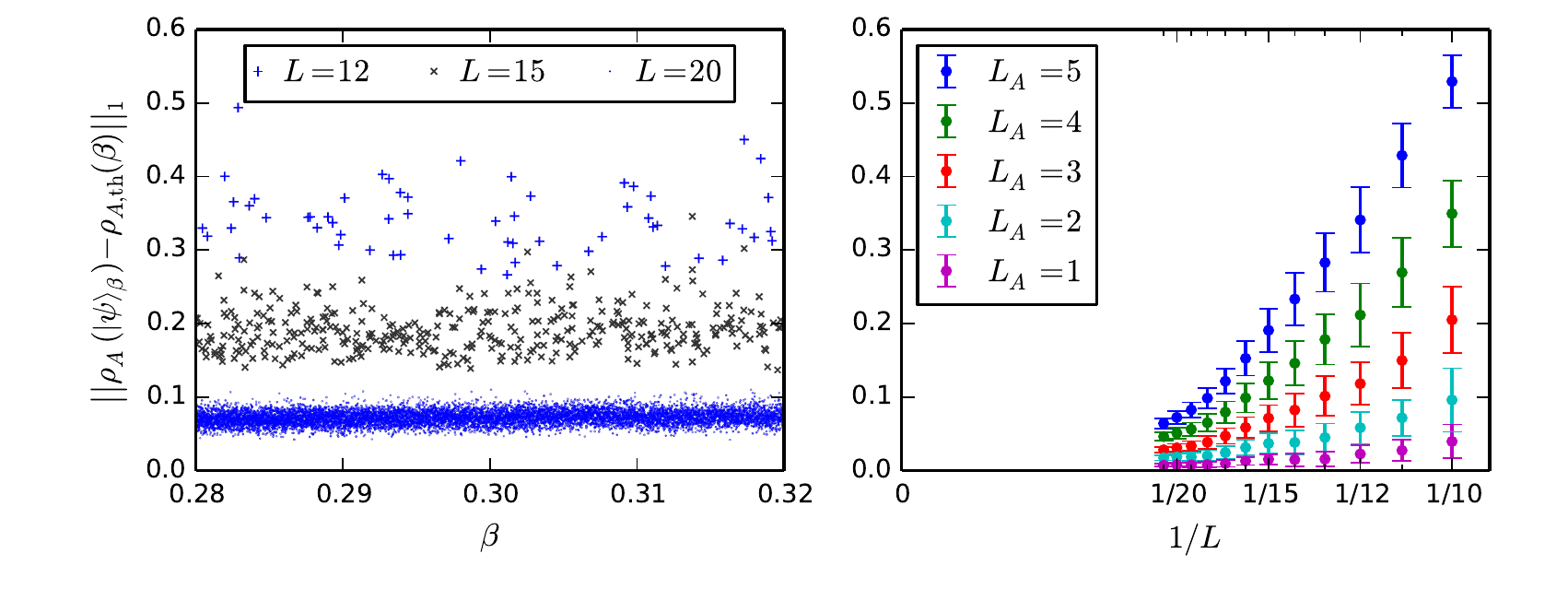}
    \caption{Trace norm distance between the canonical density matrix $\rho_{A,\mathrm{th}}(\beta)$ and the reduced density matrix $\rho_A(\ket{\psi}_\beta)$ for all eigenstates $\ket{\psi}_\beta$ in the range $0.28 < \beta < 0.32$.  The left panel plots the trace norm distance for all such eigenstates with system sizes $L=12$, $15$, and $20$, and subsystem size $L_A=5$.  The right panel plots the mean and standard deviation of the trace norm distance in this $\beta$ range for values of $L$ up to 21 and $L_A$ up to 5.}
    \label{fig:tracedist_const_La}
\end{figure*}

To quantify the extent to which Eq.\ \ref{eq:eth2} is valid, we calculate the trace norm distance $||\rho_A(|\psi\rangle_{\beta}) - \rho_{A,\mathrm{th}}(\beta)||_1$ between the reduced and canonical density matrices at various system sizes. The trace norm distance, defined as
\be 
||\rho_A(|\psi\rangle_{\beta}) - \rho_{A,\mathrm{th}}(\beta)||_1 \equiv \frac{1}{2}\tr\left[\sqrt{\left(\rho_A(|\psi\rangle_{\beta}) - \rho_{A,\mathrm{th}}(\beta)\right)^2}\right] \label{eq:tracedist}
\ee
places an upper bound on the probability difference that could result from any quantum measurement on the two density matrices \cite{wilde}.  As such, it provides an excellent measure of how distinguishable the two density matrices are.  If the trace norm distance between two finite sized density matrices is zero, they are  equal to each other element by element.

If ETH holds for all operators in subsystem A, then the results of Ref.\ \onlinecite{srednicki1998} imply that the trace norm distance should go to zero as $1/L$. The suggestion that the trace norm distance between the pure state and thermal reduced density matrices with fixed subsystem size would tend to zero was also made in Ref.\ \onlinecite{swingle}. We restrict ourselves to states in a $\beta$ range given by $0.28 < \beta < 0.32$.  In the left panel of Fig.\ \ref{fig:tracedist_const_La}, we plot the trace norm distance of every eigenstate in this $\beta$ range at $L_A=5$ for a few select system sizes.  For each system size, the distribution of the trace norm distance is nearly constant throughout the given $\beta$ range.  The right panel then takes this data for each pair of $L$ and $L_A$ and plots the mean and standard deviation of the trace norm distance against $1/L$.  The trace norm distance is tending toward zero at least linearly with $1/L$, perhaps even faster.

These results, taken together, strongly support the conjecture that ETH, as given by Eq.\ \ref{eq:eth1} holds for all operators when $V_A \ll V$.  The Schmidt eigenvalues and eigenvectors match at all energy densities, not just the energy density of the eigenstate.  Our results also imply that when $V_A \ll V$, ETH as specified by Eq.\ \ref{eq:eth2} holds.  One consequence of this is that if $V_A$ is held fixed, the density matrices $\rho_A(|\psi\rangle_\beta)$ and $\rho_{A,\mathrm{th}}(\beta)$ become elementwise equal in any basis as $V \rightarrow \infty$.

\begin{figure}[htb]
    \centering
    \begin{tabular}{c}
        \includegraphics[width=0.5\textwidth]{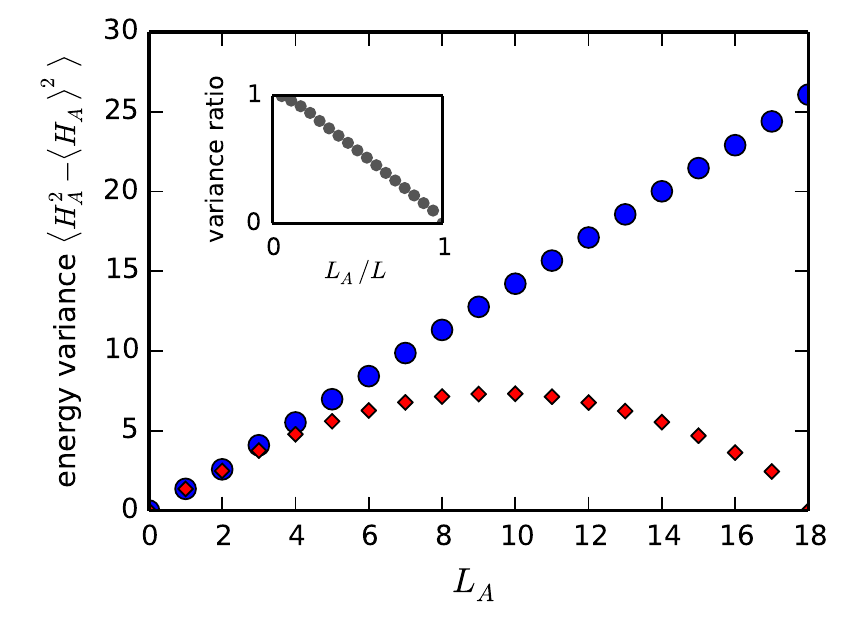} \\
        \includegraphics[width=0.5\textwidth]{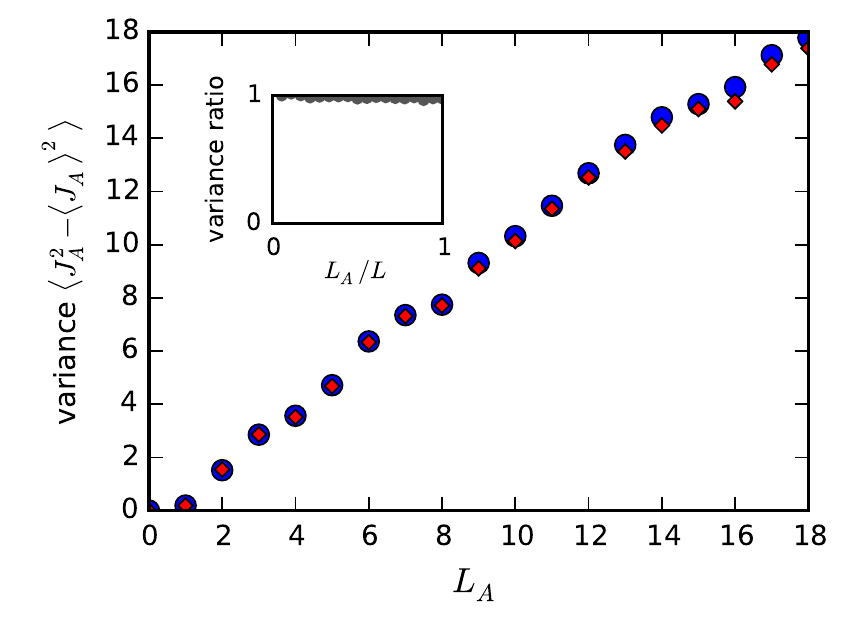}
    \end{tabular}
    \caption{(Top panel) Subsystem energy variance with respect to subsystem size $L_A$ for both the canonical ensemble (blue circular markers) and a single eigenstate $\ket{\psi}_\beta$ (red diamond markers), at $L=18$ and $\beta=0.3$.  The inset shows the ratio between the energy variances at each subsystem size, which is expected to fit $1 - L_A/L$ in the thermodynamic limit (Eq.\ \ref{eq:suppressed_variance}). (Bottom panel) The variance of an operator $J_A \equiv \sum_{i=1}^{L_A} (h_x^{(i)} \sigma_i^x + h_z^{(i)} \sigma_i^z ) + \sum_{i=1}^{L_A-1} J_z^{(i)} \sigma_i^z \sigma_{i+1}^z$, which includes the same terms as $H_A$ but does not relate to energy conservation, is plotted for comparison.  Here, the quantities $h_x^{(i)}$, $h_x^{(i)}$, and $J_z^{(i)}$ are each taken from the uniform distribution $[-1,1]$.}
    \label{fig:suppressed_variance}
\end{figure}

\begin{figure}[tbh]
    \centering
    \includegraphics[width=0.5\textwidth]{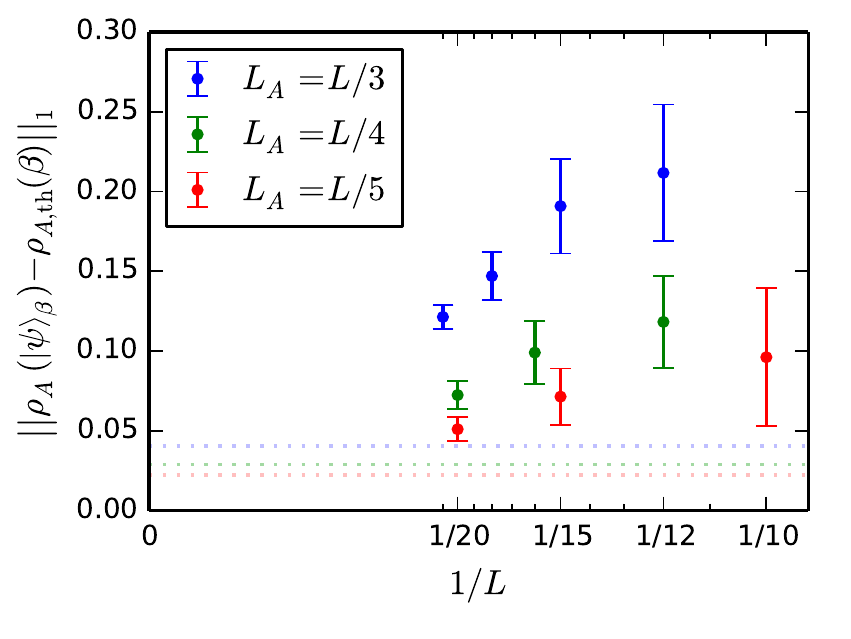}
    \caption{Trace norm distance between the canonical density matrix and reduced density matrix for constant ratio $L_A/L$ and $0.28 < \beta < 0.32$.  As in Fig.\ \ref{fig:tracedist_const_La}, the error bars represent one standard deviation away from the mean.  The horizontal lines indicate the approximate theoretical minimum each trace norm distance can take, based on the suppressed energy variance given by maximizing Eq.\ \ref{eq:tracedist_variance_bound}.}
    \label{fig:tracedist_const_ratio}
\end{figure}

\begin{figure*}[htb]
    \centering
    \begin{tabular}{cc}
        \includegraphics[width=0.5\textwidth]{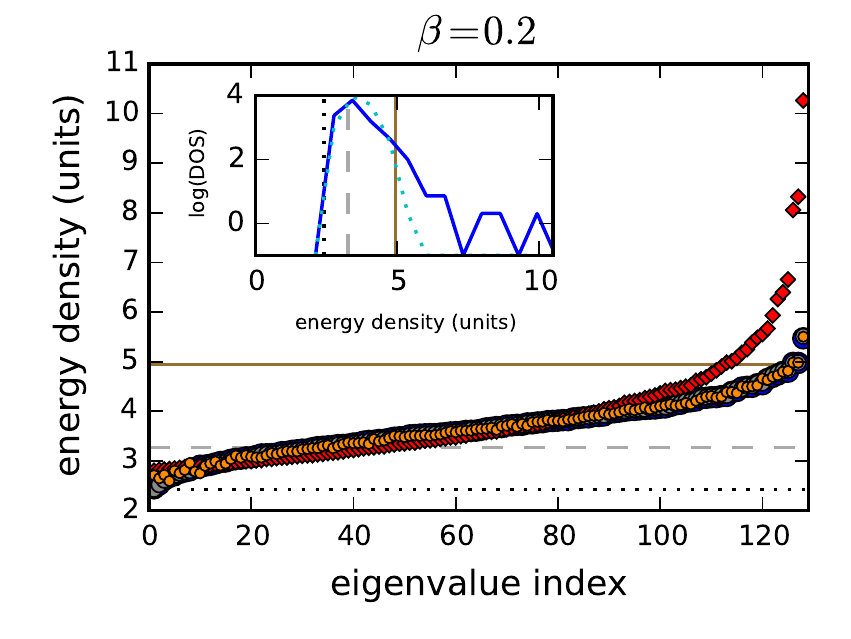} &
        \includegraphics[width=0.5\textwidth]{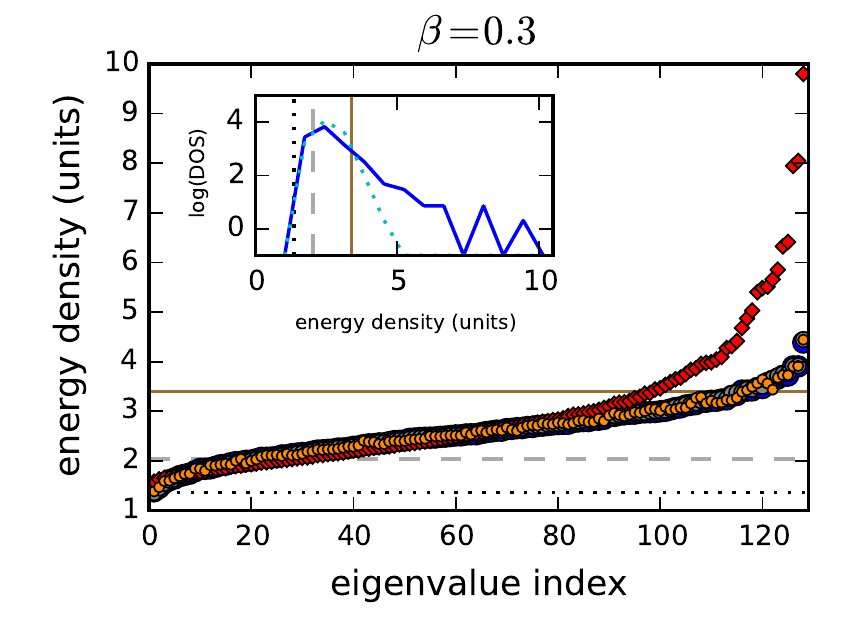} \\
        \includegraphics[width=0.5\textwidth]{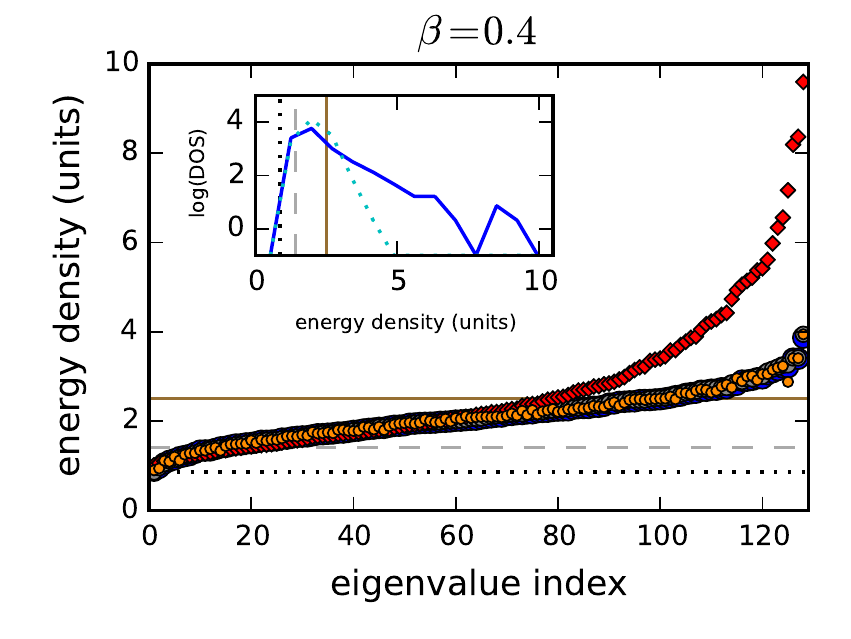} &
        \includegraphics[width=0.5\textwidth]{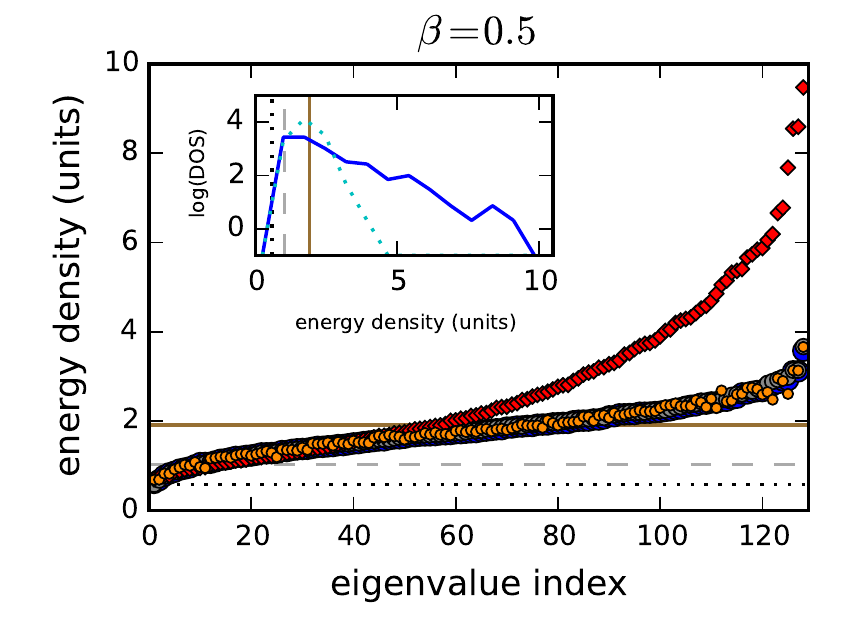} \\
    \end{tabular}
    \caption{Comparison of the four quantities defined in the inset of Fig.\ \ref{fig:spectra_overlap} for eigenstates of an $L=21$ system with $L_A=7$ at $\beta=0.2$, $0.3$, $0.4$, and $0.5$. Each inset plots a 12-bin histogram of the log of the density of states versus the energy density: the solid blue curve from a single eigenstate $\rho_A(\ket{\psi}_\beta)$ and the dotted cyan curve from the canonical ensemble $\rho_{A,\mathrm{th}}(\beta)$.  We notice that in each of the four plots, the eigenvalues of the reduced density matrix corresponding to a single eigenstate (red diamond markers) begin to deviate significantly from the other markers (in particular, the eigenvalues of the thermal reduced density matrix i.e.\ the blue markers), as the energy density reaches the critical value $e^*$ (denoted by the solid brown line), indicating breakdown of ETH beyond $e^{*}$.}
    \label{fig:econstraint}
\end{figure*}

\begin{figure*}
    \centering
    \begin{tabular}{cc}
        \includegraphics[width=0.5\textwidth]{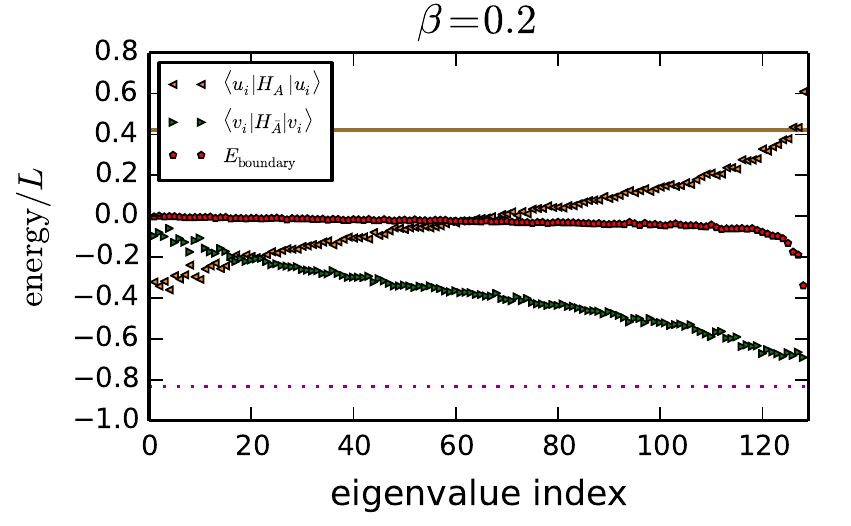} &
        \includegraphics[width=0.5\textwidth]{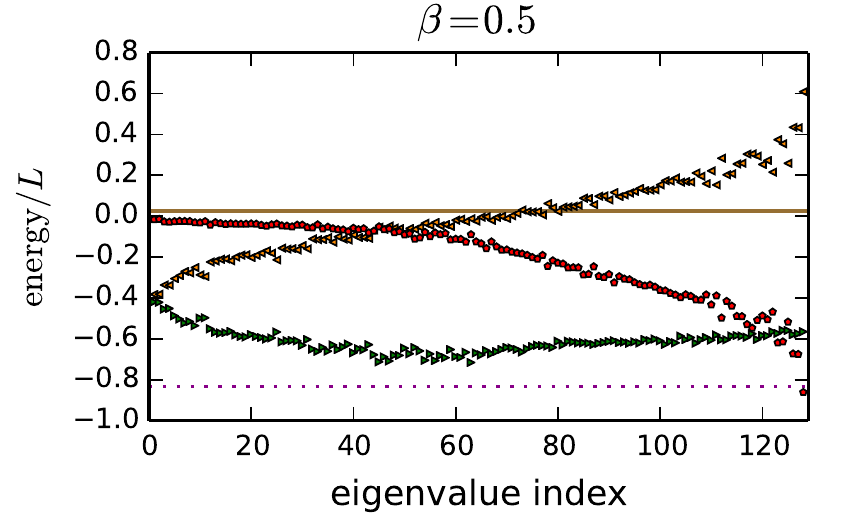}
    \end{tabular}
    \caption{Decomposition of the energy density corresponding to an eigenstate amongst the three terms in Eq.\ \ref{eq:energybalance} for $\beta=0.2$ (left panel) and $\beta=0.5$ (right panel) at $L=21$ and $L_A=7$.  The dotted magenta line marks the ground state of $H_{\overline{A}}$.  As in Fig.\ \ref{fig:spectra_overlap}, the solid brown line denotes the critical energy density $e^{*}$ for subsystem $A$.}
    \label{fig:schmidt}
\end{figure*}

\section{ETH with finite ratio $V_A/V$} \label{sec:results_finite_ratio}

In this section we will consider to what extent ETH is valid when the ratio $f \equiv V_A/V < \frac12$ is held fixed and finite as $V_A, V \rightarrow \infty$.  As demonstrated in Sec.\ \ref{sec:Snnumerical}, the von Neumann entropy of $\rho_A(\ket{\psi}_\beta)$ matches the thermal entropy in the thermodynamic limit even for finite $f < \frac12$.  In the current section we consider the extent to which other quantities match between a single eigenstate and the canonical ensemble.

There is one notable Class I operator for which ETH (in the sense of Eq.\ \ref{eq:eth1}) fails when $f$ is finite.  As mentioned in Sec.\ \ref{sec:classI}, the subsystem energy variance taken from a single eigenstate is suppressed by a factor of $(1-f)$ compared with its value in the canonical ensemble.  To demonstrate this relationship (given by Eq.\ \ref{eq:suppressed_variance}), the top panel of Fig.\ \ref{fig:suppressed_variance} shows scaling of the subsystem energy variance with subsystem size $L_A$ for both a single eigenstate and the canonical ensemble.  While the energy variance grows linearly for both for small $L_A$ in both cases, the single eigenstate energy variance has an additional term that is negative and quadratic in the subsystem size.

The bottom panel of Fig.\ \ref{fig:suppressed_variance} shows, for comparison, the variance of a different operator $J_A$ between a single eigenstate and the canonical ensemble.  The operator $J_A$ (defined in the figure's caption) is chosen to span the length of subsystem $A$ and to include the same terms as $H_A$; however, the coefficient of each term is different.  The fact that the variance of $J_A$ matches between a single eigenstate and the thermal ensemble strongly suggests that all Class I operators that do not explicitly involve energy conservation will satisfy ETH in the sense of Eq.\ \ref{eq:eth1}, even when $V_A/V$ is finite.

Let us now consider an implication of the difference in subsystem energy variance between $\rho_A(\ket{\psi}_\beta)$ and $\rho_{A,\mathrm{th}}(\beta)$.  This difference, which occurs only when $f$ is finite, suggests that the trace norm distance between the two density matrices vanishes only when $f \rightarrow 0$.  To explore this more carefully, note that the trace norm distance places a bound on the difference in expectation value of \emph{any} operator $\Lambda$ that is bounded between zero and one as
\be
\left| \tr(\rho \Lambda) - \tr(\sigma \Lambda) \right| \leq || \rho - \sigma ||_1 .
\ee
In order to calculate a lower bound on trace norm distance due to the variance difference, we must write the energy variance as a bounded operator that maximally differs between the two density matrices.  Naively one might be tempted to consider the operator $\mathcal{O}_{A,\beta}/\Delta^2 \equiv (H_A - \braket{H_A}_\beta)^2 / \Delta^2$, where for the operator to be bounded, $\Delta$ must be chosen to be the largest energy available to the system.  Since both $\mathcal{O}_{A,\beta}$ and $\Delta$ scale linearly with $V$, the expectation value of this operator is actually zero in the thermodynamic limit for both $\rho_A(|\psi\rangle_{\beta})$ and $\rho_{A,\mathrm{th}}(\beta)$.  Thus, no bound can be placed on the trace norm distance due to this particular operator.

Let us instead now consider a modified energy variance operator,
\be
\Lambda_{A,\beta,\Delta} \equiv P_{A,\beta,\Delta} \frac{\mathcal{O}_{A,\beta}}{\Delta^2}  P_{A,\beta,\Delta} ,
\ee
where $\Delta$ is an arbitrary energy scale and $P_{A,\beta,\Delta}$ projects onto the subspace where $\mathcal{O}_{A,\beta}/\Delta^2$ has eigenvalues in the range $[0,1]$, thus making $\Lambda_{A,\beta,\Delta}$ a bounded operator.  This operator considers the energy variance within a restricted window of width $2\Delta$ about the mean energy.

To arrive at an approximate bound due to this operator, let us assume that the energy histograms of $\rho_{A,\mathrm{th}}(\beta)$ and $\rho_A(|\psi\rangle_{\beta})$ are given by normal distributions with variance $\sigma_\mathrm{th}^2$ and $\sigma_\psi^2 = (1-f) \sigma_\mathrm{th}^2$, respectively.  Since both distributions have the same mean, the difference in expectation values is expected to be
\begin{align} \label{eq:tracedist_variance_bound}
D &\equiv \tr [ \rho_{A,\mathrm{th}}(\beta) \Lambda_{A,\beta,\Delta} ] - \tr [ \rho_A(\ket{\psi}_\beta) \Lambda_{A,\beta,\Delta} ] \notag \\
  &= \frac{1}{\Delta^2} \int_{-\Delta}^{\Delta} \left[ \frac{e^{-E^2/2\sigma_\mathrm{th}^2}}{\sigma_\mathrm{th}\sqrt{2\pi}} - \frac{e^{-E^2/2\sigma_\psi^2}}{\sigma_\psi\sqrt{2\pi}} \right] E^2\,dE .
\end{align}
Given $\sigma_\mathrm{th}$ and $f$, it is possible to find $\Delta$ numerically such that $D$ is maximized.  Although $\Delta$ is proportional to $\sqrt{V}$, the value of $D$ itself is independent of $V$ as $V \rightarrow \infty$, since $\sigma_\mathrm{th}$ also scales with $\sqrt{V}$.  The maximum quantity $D$ then provides a lower bound on the trace norm distance between $\rho_{A,\mathrm{th}}(\beta)$ and $\rho_A(|\psi\rangle_{\beta})$ in the thermodynamic limit\cite{footnote:higher_moments}.

Let us now turn to our results on the scaling of trace norm distance with system size when the ratio $f = L_A/L$ is held fixed as $L, L_A \rightarrow \infty$, which are shown in Fig.\ \ref{fig:tracedist_const_ratio}.  Although there are few points available for each ratio, the trend is clearly for the trace norm distance to decrease as $L$ increases.  The horizontal, dotted lines denote the theoretical minimum each trace norm distance can taken, given by Eq.\ \ref{eq:tracedist_variance_bound}.  Remarkably, for each subsystem ratio, the trace norm distance rapidly approaches this lower bound, suggesting that the bound may actually provide the result in the thermodynamic limit.  This in turn implies that other operators which do not involve energy conservation are likely to have equal expectation values for a single eigenstate and for the canonical ensemble.

We now turn to results on the entanglement spectrum when $f$ is a significant fraction of the total system size.  As discussed in Sec.\ \ref{sec:classII}, if the constraint in Eq.\ \ref{eq:econstraint} is violated, the entanglement spectrum cannot match above a critical energy density $e^{*} = e/f$ (see Eq.\ \ref{eq:ecrit}), where $e$ is the energy density of the state $|\psi\rangle_\beta$.  Fig.\ \ref{fig:econstraint} shows the comparison of spectra of four different quantities considered in Sec.\ \ref{sec:extracting} for several different energy densities of the reference state $|\psi\rangle_\beta$ with $f = 1/3$, at four different values of $\beta$. With $f=1/3$, the energy constraint Eq.\ \ref{eq:econstraint} is violated, and therefore we expect that the entanglement spectrum should deviate from the actual spectrum of the Hamiltonian at least beyond the critical energy density $e^{*} = e/f$.  We find for each value of $\beta$ that significant deviation starts to occur essentially right at this critical energy density.

Surprisingly, even though the entanglement spectrum does not match the actual spectrum beyond the energy density $e^{*}$, the expectation values  $\langle  u_i|H_A| u_i\rangle/L_A$ continue to match the energy eigenvalues of the actual Hamiltonian! 
To understand this phenomenon better, we analyze the different terms in Eq.\ \ref{eq:energybalance}. As argued in Sec.\ \ref{sec:classII}, the only way $\langle  u_i|H_A| u_i\rangle$ can exceed the total energy $E$ of the eigenstate is, if the $E_\mathrm{boundary}$ term,

\be
E_\mathrm{boundary} \equiv \sum_{j} \sqrt{\frac{\lambda_j}{\lambda_{i0}}} \langle u_{i0}| \otimes \langle v_{i0}|H_{A\Abar}| u_j\rangle \otimes |v_j\rangle,
\ee
scales with the total system size.  We find that this is indeed the case, as shown in Fig.\ \ref{fig:schmidt}. In agreement with the general considerations in Sec.\ \ref{sec:classII}, the Schmidt eigenvalues deviate from their ETH predicted values beyond $e^{*}$ (Fig.\ \ref{fig:econstraint}) and become considerably smaller.

To summarize the results of this section, we provided evidence that ETH holds for all Class I operators not related to energy conservation.  For Class II operators, there is a critical energy density $e^*$ beyond which ETH definitely fails, even though, surprisingly, the eigenvectors still seem to be correct. (We leave the understanding of this result for future studies.)

\begin{figure*}[htb]
    \centering
    \begin{tabular}{cc}
        \includegraphics[width=0.5\textwidth]{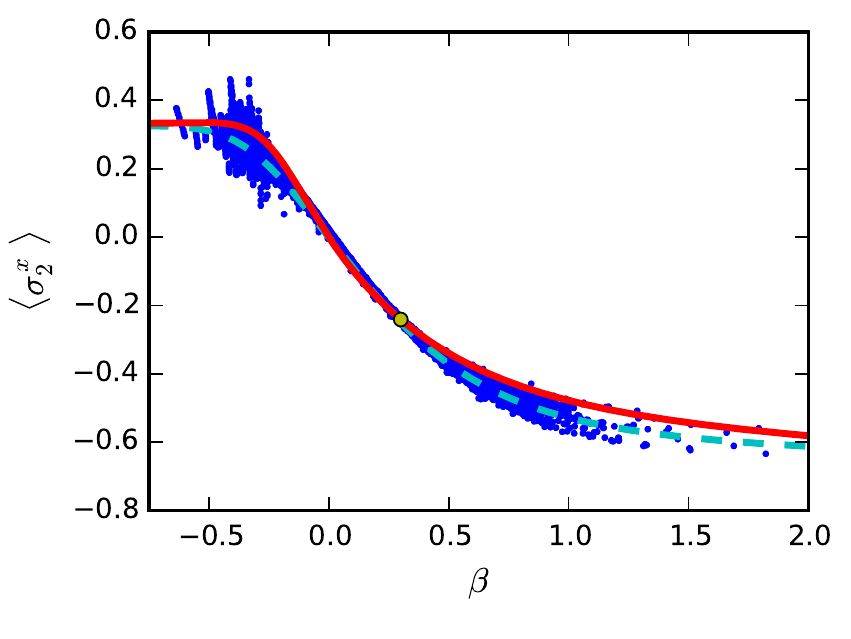} &
        \includegraphics[width=0.5\textwidth]{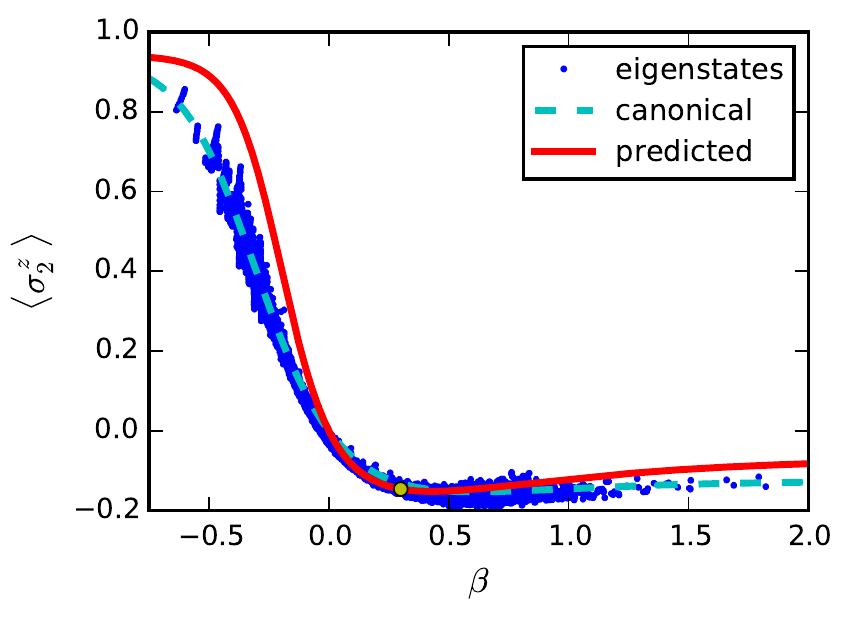} \\
        \includegraphics[width=0.5\textwidth]{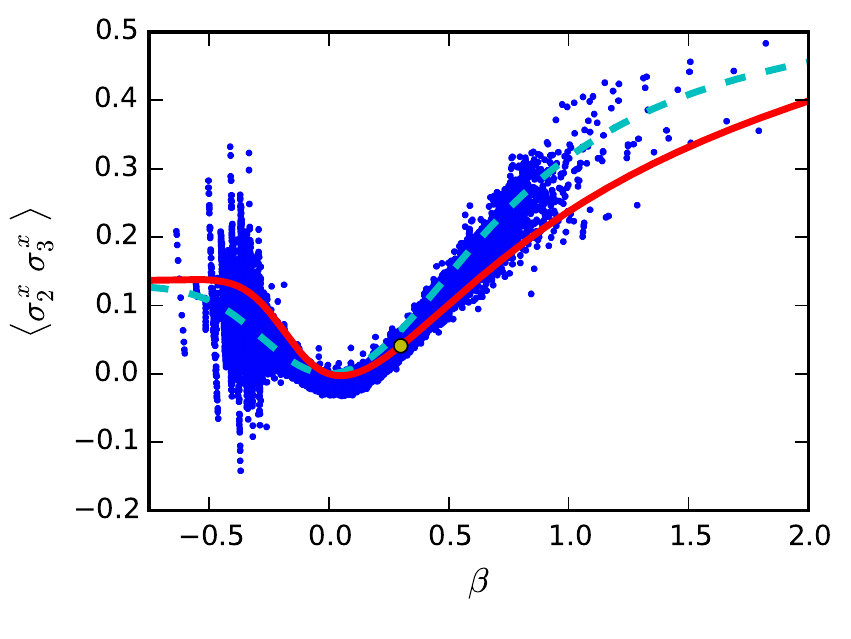} &
        \includegraphics[width=0.5\textwidth]{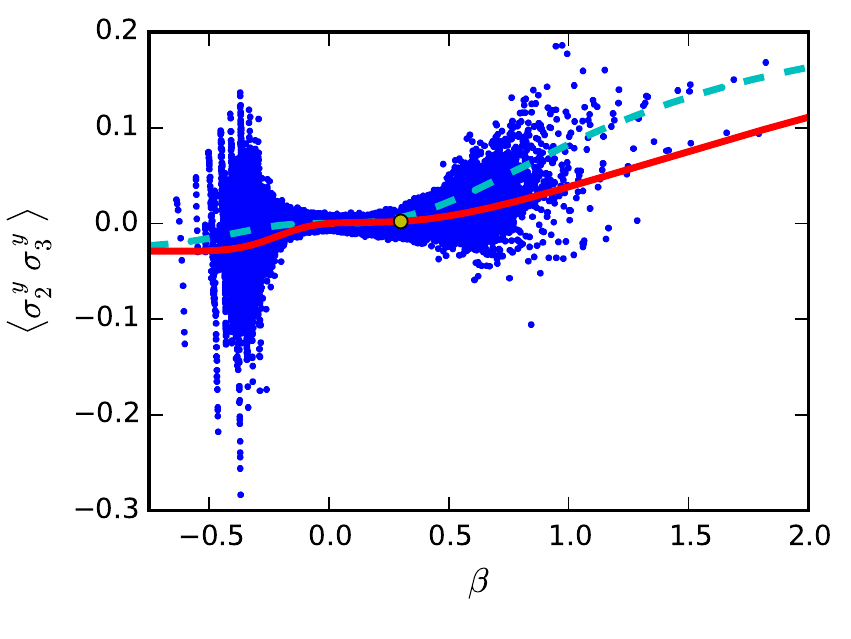} \\
    \end{tabular}
    \caption{Equal time correlators for an $L=21$ system plotted against inverse temperature $\beta$.  The blue dots denote the expectation value with respect to each eigenstate, the dashed cyan curve plots the expectation value in the canonical ensemble, and the red curve plots the expectation value predicted from a single eigenstate at $\beta_0 = 0.3$ (yellow dot) by raising the $L_A=4$ density matrix to the power $\beta/\beta_0$ and rescaling it to have unit trace.}
    \label{fig:equal_time_correlators}
\end{figure*}

\section{An application: equal-time correlators as a function of temperature from a single eigenstate}  \label{sec:corr}
In the previous sections we provided evidence that a single eigenstate encodes the full Hamiltonian. As an application of this result, we now calculate correlation functions at arbitrary temperatures using a \textit{single eigenstate} $|\psi\rangle_\beta$. The basic idea is similar to the relation between the Renyi entropies and the free energy densities (Eq.\ \ref{eq:Sn}).

In particular, consider the correlation function, 
\be 
\langle O(x) O(y) \rangle_{\beta,n} = \frac{\tr_A \left(\rho^n_A(|\psi\rangle_{\beta}) O(x) O(y)\right)}{\tr_A \left(\rho^n_A(|\psi\rangle_{\beta})\right)} \label{eq:corr1}
\ee
where $x,y$ are located in subsystem $A$, away from the boundary. Using Eqs.\ \ref{eq:eth2}, \ref{eq:eth3} to leading order in the subsystem size, $\langle O(x) O(y) \rangle_{\beta,n}$ equals the expectation value of the operator $O(x) O(y)$ at a temperature $(n\beta)^{-1}$. 

Fig.\ \ref{fig:equal_time_correlators} shows the expectation values of local operators within subsystem $A$ as a function of $\beta$, as predicted from a single eigenstate at inverse temperature $\beta_0$ (indicated by a yellow dot on the red curve). We choose operators that are as far away from the subsystem boundary as possible so as to minimize the finite size corrections.  Even though the agreement with the canonical ensemble is not perfect, the qualitative trends and the numerical values match incredibly well, given the modest total system sizes to which we are restricted.  These predicted correlators also undoubtedly suffer from corrections expected due to the conical singularity at the boundary of $A$ in Eq.\ \ref{eq:corr1}.

\section{Summary and discussion} \label{sec:discuss}
In this paper, we analyzed the  structure of reduced density matrices corresponding to the eigenstates of generic, non-integrable quantum systems. We argued that given an eigenstate $|\psi\rangle_\beta$ with energy density $e$ and a corresponding temperature $\beta^{-1}$, the reduced density matrix for a subsystem $A$ is given by

\be 
\rho_A(|\psi\rangle_{\beta}) = \rho_{A,\mathrm{th}}(\beta) \nonumber
\ee
where  

\be 
\rho_{A,\mathrm{th}}(\beta) =  \frac{\tr_{\overline{A}} \left( e^{-\beta H}\right)}{\tr \left(e^{-\beta H}\right)} \nonumber
\ee
if the condition $V_A \ll V$ is satisfied.  This means that for a fixed eigenstate $|\psi\rangle_\beta$, one can always extract the properties of the corresponding Hamiltonian at arbitrary energy densities by taking $V_A/V \rightarrow 0$ as the limits $V_A, V \rightarrow \infty$ are taken.  Remarkably, even when $V_A/V$ ($< 1/2$) is taken to be fixed and finite, one can still access many properties of the underlying Hamiltonian for a range of energy densities in the interval described in Eq.\ \ref{eq:ecrit}.

We also introduced the notion of ``equithermal'' (Class I) and ``non-equithermal'' (Class II) operators. In a canonical ensemble at temperature $T$, the expectation value of Class I operators depends only on the properties of the underlying Hamiltonian at temperature $T$, while the same is not true for Class II operators. Our results strongly suggest that all Class I operators not involving energy conservation satisfy Eq.\ \ref{eq:eth1} as long as $V_A < V/2$. For Class II operators not related to energy conservation, Eq.\ \ref{eq:eth1} is seemingly again satisfied as long as the constraint in Eq.\ \ref{eq:econstraint} holds.

We also provided analytical results for the Renyi and von Neumann entropies of infinite temperature eigenstates of a particle number conserving model. These results substantiate our numerical results for the energy-only conserving model. In particular, we find that the von Neumann entanglement entropy for a subsystem of size $V_A$ equals the thermal entropy for that subsystem as long as $V_A < V/2$, and therefore follows the so called `Page curve' \cite{page1993} at all non-zero temperatures, thus generalizing the original work of Page and others \cite{page1993}, and in agreement with the recent work on large central charge CFTs \cite{asplund2014, caputa2014, kaplan2015}.

In this paper we only considered contiguous subsystems. It seems reasonable to conjecture that Eq.\ \ref{eq:eth2} continues to hold as long as the support of operator $O$ can be chosen to lie in a subsystem which is not necessarily contiguous and whose volume satisfies $V_A \ll V$. This encompasses the expectation values of local operators such as $\langle O(\vec{x}) O(0)\rangle$, where $O(\vec{x})$ is localized at point $\vec{x}$ and $|\vec{x}|$ can be greater than $L/2$ (where $L$ is the linear dimension of the total system).

Let us mention some of the practical implications of our results. Firstly, the fact that a single eigenstate encodes properties of the full Hamiltonian could potentially be a useful numerical tool. For example, one could imagine targeting a finite energy density eigenstate of a Hamiltonian $H$  by variationally minimizing the energy of the Hamiltonian $(H-E)^2$ with respect to trial wavefunctions. The techniques in this paper would then allow one to access thermal properties of the Hamiltonian without  directly calculating the partition function, which could be extremely helpful for Hamiltonians that suffer from the sign problem.

Secondly, owing to the recent progress in single atom imaging techniques in cold atomic systems \cite{greiner2009}, one can now access non-local operators experimentally \cite{endres, bloch, daley2012, greiner2015}. This potentially allows one to check some of our predictions pertaining to the violation of ETH in cold atomic systems. For example, one can perform a quantum quench on a low entanglement state which would at sufficiently long times lead to a thermal state in the same sense as Eq.\ \ref{eq:eth2}. This in principle allows one to determine the underlying Hamiltonian of a cold atomic system by performing tomography on a small subsystem to obtain the corresponding reduced density matrix, and then taking its logarithm.

We conclude by posing a few questions and future directions.

In this paper we extracted equal-time correlators at different temperatures using a single eigenstate. It will be interesting to see if a similar method also works for unequal time correlators at arbitrary temperatures. The main difference is that this requires calculating expressions such as Eq.\ \ref{eq:corr1} at an \textit{imaginary} exponent, and estimating the effects due to the conical singularity  in this case requires further study.

As mentioned above, we expect that all our discussion carries over to time-evolved states  as well since such states are expected to also have thermal behavior at long times in the same sense as a single finite energy density eigenstate. If so, does the time scale for thermalization for a given operator (i.e.\ the time it takes for the expectation value of the operator to become equal to its canonical expectation value) depend on whether the operator is Class I (equithermal) or Class II (non-equithermal)?

Another question concerns the subleading corrections to the entanglement entropy. One expects that there always exist subleading area-law contributions to the entanglement entropy (either von Neumann or Renyi) of a single eigenstate. Are these contributions also captured correctly in the entanglement entropies calculated via a thermal reduced density matrix? Perhaps a more interesting question is whether the mutual information of two disjoint intervals (which cancels out both the volume law contribution and the area law contribution) takes the same value for a single eigenstate and its canonical counterpart.

We are grateful to Leon Balents, John Cardy, Matthew Fisher, Tom Hartman, Patrick Hayden, Pavan Hosur, Max Metlitski, Olexei Motrunich, Chetan Nayak, Anatoli Polkovnikov,  Marcos Rigol, Lea Santos, Brian Swingle, and especially David Huse and Xiaoliang Qi  for conversations about this work.  This research was supported in part by the National Science Foundation, under Grant No.\ DMR-14-04230 (JRG), by the Caltech Institute of Quantum Information and Matter, an NSF Physics Frontiers Center with support of the Gordon and Betty Moore Foundation (JRG), and by the Gordon and Betty Moore Foundations' EPiQS Initiative through Grant GBMF4304 (TG). This research was also supported in part by the National Science Foundation under Grant No.\ NSF PHY11-25915.  We also acknowledge support from the Center for Scientific Computing at the CNSI and MRL: an NSF MRSEC (DMR-1121053) and NSF CNS-0960316.


\begin{thebibliography}{x}

\bibitem{bisognano1976} J. J. Bisognano and E. H. Wichmann,  J. Math. Phys. 17, 303 (1976); J. J. Bisognano and E. H. Wichmann, J. Math. Phys. 16, 985 (1975).

\bibitem{susskind2004} Leonard Susskind, \textit{An Introduction To Black Holes, Information And The String Theory Revolution: The Holographic Universe}, World Scientific Publishing Company (2004).

\bibitem{holzhey1994} C. Holzhey, F. Larsen and F. Wilczek, Nucl. Phys. B 424, 443 (1994).

\bibitem{callan1994} C. G. Callan and F. Wilczek, Phys. Lett. B 333, 55 (1994).

\bibitem{calabrese2004} P. Calabrese and J. Cardy, J. Stat. Mech. 0406:P06002 (2004).

\bibitem{casini2011}  Horacio Casini, Marina Huerta, and Robert C. Myers, JHEP 1105:036 (2011).

\bibitem{levin2006} M. Levin and X. G. Wen, Phys. Rev. Lett. 96, 110405 (2006).

\bibitem{kitaev2006} A. Kitaev and J. Preskill, Phys. Rev. Lett. 96,
110404 (2006).

\bibitem{li2008} H. Li and F. D. M. Haldane,  Phys. Rev. Lett. 101, 010504 (2008).

\bibitem{zhang2012} Yi Zhang, Tarun Grover, Ari Turner, Masaki Oshikawa, and Ashvin Vishwanath, Phys. Rev. B 85, 235151 (2012).

\bibitem{bombelli1986} L. Bombelli, R. K. Koul, J. Lee, and R. D. Sorkin, Phys. Rev. D 34, 373 (1986).

\bibitem{srednicki1993} M. Srednicki, Phys. Rev. Lett. 71, 666 (1993). 

\bibitem{srednicki1994} M. Srednicki, Phys. Rev. E 50, 888 (1994); J. Phys. A 29, L75 (1996); J. Phys. A 32, 1163 (1999).

\bibitem{deutsch1991} J. M. Deutsch, Phys. Rev. A 43, 2046 (1991).

\bibitem{srednicki1998} M. Srednicki, J. Phys. A: Mathematical and General 32, 1163 (1998).

\bibitem{gefen1997} B. L. Altshuler, Y. Gefen, A. Kamenev, and L. S. Levitov, Phys.Rev.Lett. 78, 2803 (1997).

\bibitem{gornyi2005} I. V. Gornyi, A. D. Mirlin, and D. G. Polyakov, Phys. Rev. Lett. 95, 206603 (2005).

\bibitem{basko2006} D. Basko, I. Aleiner, and B. Altshuler, Annals of Physics 321, 1126 (2006).

\bibitem{huse2007} V. Oganesyan and D. Huse, Physical Review B 75, 155111 (2007).

\bibitem{bela} B. Bauer and C. Nayak,  J. Stat. Mech. (2013) P09005.

\bibitem{imbrie}  J. Z. Imbrie, arXiv:1403.7837.

\bibitem{nandkishore2014} Rahul Nandkishore and David A. Huse, Annual Review of Condensed Matter Physics 6, 15 (2015).

\bibitem{kagan1} Y. Kagan and L. A. Maksimov, J. Phys. C 7, 2791 (1974).

\bibitem{kagan2} Y. Kagan and L. A. Maksimov, Zh. Eksp. Teor. Fiz. 87, 348 (1984).

\bibitem{grover2014} T. Grover and M. P. A. Fisher, J. Stat. Mech. 10, 10010 (2014).

\bibitem{muller_aps} M. Schiulaz and M. M\"uller, AIP Conf. Proc. 1610, 11 (2014).

\bibitem{muller2014} M. Schiulaz, A. Silva, and M. M\"uller, Phys. Rev. B 91, 184202 (2015).

\bibitem{hickey} J. M. Hickey, S. Genway, and J. P. Garrahan, arXiv:1405.5780.

\bibitem{yao2014} N. Y. Yao, C. R. Laumann, J. I. Cirac, M. D. Lukin, and J. E. Moore, arXiv:1410.7407 (2014).

\bibitem{roeck1} W. De Roeck and F. Huveneers, Comm. Math. Phys. 332, 1017 (2014).

\bibitem{roeck2} W. De Roeck and F. Huveneers, Phys. Rev. B 90, 165137 (2014).

\bibitem{roeck3} W. De Roeck and F. Huveneers, arXiv:1409.8054. 

\bibitem{abanin2015} Z. Papic, E. M. Stoudenmire, and Dmitry A. Abanin,  arXiv:1501.00477.

\bibitem{rigol2008} M. Rigol, V. Dunjko, and M. Olshanii, Nature 452, 854 (2008).

\bibitem{lauchli2011} G. Biroli, C. Kollath, and A. L\"auchli, Phys. Rev. Lett. 105, 250401 (2010).

\bibitem{srednicki2011} Marcos Rigol and Mark Srednicki, Phys. Rev. Lett. 108, 110601 (2012).

\bibitem{hosur} Pavan Hosur and Xiao-Liang Qi, arXiv:1507.04003.

\bibitem{bloch} Michael Schreiber, Sean S. Hodgman, Pranjal Bordia, Henrik P. Luschen, Mark H. Fischer, Ronen Vosk, Ehud Altman, Ulrich Schneider, and Immanuel Bloch, Science 349, 842 (2015).

\bibitem{greiner2009} Waseem S. Bakr, Jonathon I. Gillen, Amy Peng, Simon Folling, and Markus Greiner, Nature 462, 74 (2009).

\bibitem{deutsch2010} J. M. Deutsch, New J. Phys. 12, 075021 (2010).

\bibitem{lauchli2013} Lars Bonnes, Hannes Pichler, and Andreas M. Lauchli, Phys. Rev. B 88, 155103 (2013).

\bibitem{melko2011} Rajiv R. P. Singh, Matthew B. Hastings, Ann B. Kallin, and Roger G. Melko, Phys. Rev. Lett. 106, 135701 (2011).

\bibitem{santos2012} Lea F. Santos, Anatoli Polkovnikov, and Marcos Rigol, Phys. Rev. E 86, 010102(R) (2012).

\bibitem{storms2014} Michelle Storms, Rajiv R. P. Singh, Phys. Rev. E 89, 012125 (2014).

\bibitem{lai2014} Hsin-Hua Lai and Kun Yang, Phys. Rev. B 91, 081110 (2015).

\bibitem{pathria} R. K. Pathria, \textit{Statistical Mechanics}, Butterworth-Heinemann (1996).

\bibitem{baez} John C. Baez, arXiv:1102.2098.

\bibitem{cardy2014} John Cardy, Phys. Rev. Lett. 112, 220401 (2014).

\bibitem{asplund2014} Curtis T. Asplund, Alice Bernamonti, Federico Galli, and Thomas Hartman, JHEP 2015:171 (2015).

\bibitem{caputa2014} Pawel Caputa, Joan Simon, Andrius Stikonas, and Tadashi Takayanagi, JHEP 2015:102 (2015).

\bibitem{kaplan2015} A. Liam Fitzpatrick, Jared Kaplan, and Matthew T. Walters, arXiv:1501.05315.

\bibitem{lubkin1978} E. Lubkin, J. Math. Phys. 19, 1028 (1978).

\bibitem{lloyd1988} S. Lloyd and H. Pagels, Ann. Phys., NY, 188, 186 (1988).

\bibitem{page1993} D. N. Page, Phys. Rev. Lett. 71, 1291 (1993); Foong S. K. and S. Kanno, Phys. Rev. Lett. 72, 1148 (1994); J. Sanchez-Ruiz, Phys. Rev. E 52, 5653 (1995); S. Sen, Phys. Rev. Lett. 77, 1 (1996).

\bibitem{footnote:david} We are grateful to David Huse to pointing this out to us.

\bibitem{rigol2014} M. Rigol, Phys. Rev. Lett. 112, 170601 (2014).

\bibitem{tasaki98} H. Tasaki, Phys. Rev. Lett. 80, 1373 (1998).

\bibitem{goldstein2006} S. Goldstein, J.L. Lebowitz, R. Tumulka, and N. Zangh`i, Phys. Rev. Lett. 96, 050403 (2006).

\bibitem{popescu2006} S. Popescu, A.J. Short, and A. Winter, Nature Phys. 2, 754 (2006).

\bibitem{santos2010} Lea F. Santos and Marcos Rigol, Phys. Rev. E 81, 036206 (2010).

\bibitem{deutsch2013} J. M. Deutsch, Haibin Li, and Auditya Sharma, Phys. Rev. E 87, 042135 (2013).

\bibitem{kim2014} Hyungwon Kim, Tatsuhiko N. Ikeda, and David A. Huse, Phys. Rev. E 90, 052105 (2014).

\bibitem{calabrese2010} Pasquale Calabrese, Massimo Campostrini, Fabian Essler, and Bernard Nienhuis, Phys. Rev. Lett. 104, 095701 (2010).
 
\bibitem{liu2015} Anatoly Dymarsky, Hong Liu, arXiv:1511.06680.

\bibitem{wilde}  See e.g.\ M. M. Wilde, \textit{Quantum Information Theory}, Cambridge University Press (2013).

\bibitem{swingle} Brian Swingle and Isaac H. Kim, Phys. Rev. Lett. 113, 260501 (2014).

\bibitem{footnote:higher_moments} One could also consider bounds on the trace norm distance due to higher moments $\langle (H_A - \langle H_A \rangle)^n \rangle$, but the most stringent bound comes from $n=2$.

\bibitem{endres} M. Endres, M. Cheneau, T. Fukuhara, C. Weitenberg, P. Schauss, C. Gross, L. Mazza, M.C. Banuls, L. Pollet, I. Bloch, and S. Kuhr, Science 334, 200 (2011).

\bibitem{daley2012} A. J. Daley, H. Pichler, J. Schachenmayer, and P. Zoller, Phys. Rev. Lett. 109, 020505 (2012).

\bibitem{greiner2015} Rajibul Islam, Ruichao Ma, Philipp M. Preiss, M. Eric Tai, Alexander Lukin, Matthew Rispoli, and Markus Greiner, arXiv:1509.01160.

\end{thebibliography}
\end{document}